\documentclass[fleqn,usenatbib]{mnras}

\usepackage[T1]{fontenc}

\DeclareRobustCommand{\VAN}[3]{#2}
\let\VANthebibliography\thebibliography
\def\thebibliography{\DeclareRobustCommand{\VAN}[3]{##3}\VANthebibliography}

\usepackage{graphicx}	
\usepackage{amsmath}	
\usepackage{amssymb}	

\usepackage{tablefootnote}

\usepackage{newtxtext}
\usepackage[varg]{newtxmath}

\title[Unveiling stellar aurorae]{Unveiling stellar aurorae: Simulating auroral emission lines in hot stars induced by high-energy irradiation}

\author[M. Kajan, J. Krtička and J. Kubát]{
Michal Kajan$^{1}$\thanks{E-mail: kajan@mail.muni.cz}, Jiří Krtička$^{1}$, Jiří Kubát $^{2}$
\\
$^{1}$Department of Theoretical Physics and Astrophysics, Faculty of Science, Masaryk University, Kotlářská 2, Brno, Czech Republic \\
$^{2}$Astronomical Institute, Academy of Sciences of the Czech Republic, Fri\v{c}ova 298, 251 65 Ond\v{r}ejov, Czech Republic
}

\date{Accepted XXX. Received YYY; in original form ZZZ}

\pubyear{2023}

\begin{document}
\label{firstpage}
\pagerange{\pageref{firstpage}--\pageref{lastpage}}
\maketitle

\begin{abstract}
Auroral emission lines result from the interaction between magnetic field and stellar wind, offering valuable insights into physical properties and processes occurring within magnetospheres of celestial bodies. 
While extensively studied in planetary and exoplanetary atmospheres, in ultra-cool dwarfs, and as radio emission from early-type stars, the presence of specific auroral emission lines in hot star spectra remains unexplored.
In this study, we utilized TLUSTY code to simulate the auroral lines, while modelling the effect of the interaction between stellar wind and magnetosphere through X-ray irradiation.
Utilizing high-resolution synthetic spectra generated from model atmospheres, we identified potential candidate lines indicative of auroral emission, which were absent in non-irradiated spectra.
Emission lines in synthetic spectra were present primarily in the infrared domain.
The most prominent line generated by irradiation was \ion{He}{ii} 69458\,\AA, which appeared in all our model atmospheres with effective temperatures ranging from 15\,kK to 30\,kK.
We also calculated the minimum irradiation required to detect emission in this most prominent line.
The presence of emission lines was interpreted by considering changes in the population of different excited states of given atoms.
Besides the appearance of infrared emission lines, high-energy irradiation causes infrared excess.
To complement our simulations, we also searched for auroral lines in Far Ultraviolet Spectroscopic Explorer (FUSE) observations, which are deposited in the Multimission Archive at Space Telescope (MAST) catalogue.
The comparison of observed spectra with synthetic spectra did not identify any possible candidate emission lines in FUSE spectra.

\end{abstract}


\section{Introduction}

Significant populations of A and B-type main sequence stars exhibit stable, large-scale magnetic fields that are detectable from spectropolarimetry.
These magnetic fields typically possess a surface strength on the order of a few kilogauss 
\citep{2015IAUS..307..342M,2018MNRAS.475.5144S,2020AzAJ...15a..93Y}.
The majority of these magnetic stars display a dominant dipole magnetic topology, although few exceptions exist. 
A model known as the Rigidly Rotating Magnetosphere (RRM), proposed by \cite{2005MNRAS.357..251T}, describes many of the observed features associated with circumstellar magnetospheres of magnetic stars.
According to this model, the matter accumulates in magnetospheric clouds corresponding to the minima of the effective potential along each field line.

The magnetosphere is filled with the stellar wind confined by the strong magnetic field.
Various parameters can affect attributes and dynamics of the magnetosphere, including mass loss, terminal velocity of the wind, and surface magnetic field strength as the most influential.
For quantitive effects of the magnetic field, \cite{2002ApJ...576..413U} and \cite{2008MNRAS.385...97U} introduced a parameter for magnetic confinement $\eta_*$.
If $\eta_*>1$, the wind is magnetically confined, which means that closed magnetic field lines exist in the magnetosphere.

The magnetic confinement parameter is associated with the Alfvén radius $R_\text{A}$, the radius where magnetic field energy density is equal to wind kinetic energy density.
The position at which the centrifugal force, in a frame rigidly rotating with the star, balances the gravitational force, is denoted as the Kepler corotation radius, $R_\text{K}$.
The comparison of these radii defines two distinct types of magnetospheres.
In summary, the dynamical magnetosphere is defined as $R_\text{A}<R_\text{K}$ and the centrifugal magnetosphere is defined as $R_\text{A}>R_\text{K}$.
Moreover, the centrifugal magnetosphere contains regions where the trapped material can corotate \citep[see][for a review]{2016smfu.book..347R}.
Numerous observational effects exist as a piece of evidence for the magnetosphere in various spectral domains, specifically X-ray domain \citep[][]{2015MNRAS.452.2641N},
 ultraviolet \citep[UV,][]{2013MNRAS.431.2253M}, near-infrared \citep{2015A&A...578A.112O},
radio \citep{2021MNRAS.507.1979L}, and H$\alpha$ \citep{2020MNRAS.499.5366O}.
Further details, including a visual schematic of the stellar magnetosphere, are provided by \cite{2020pase.conf...54S}.

Focusing on the most energetic part of the spectrum, a significant fraction of high energy emissions in magnetic hot stars comes from magnetically confined wind shocks (MCWS). 
This phenomenon happens as the magnetically confined wind from different magnetic poles collides near the magnetic equator and creates wind shocks \citep{2022hxga.book...46U}.
The brightness temperature of MCWS can be of the order of $\sim 10^7-10^8$\,K \citep{2016AdSpR..58..680U}.

Furthermore, in the radio part of the spectrum of a few magnetic stars, we can observe a curious effect created by Electron Cyclotron Maser Emission (ECME).
This emission is non-common mainly because of its high polarization \citep{2004A&A...418..593T}.
The ECME is supposed to originate in the centrifugal magnetosphere, where fast electrons are trapped and accelerated towards the star while they emit radio emissions. 
Fast electrons are likely produced during reconnection events associated with matter leakage from the magnetospheres \citep{2022MNRAS.513.1449O}.

The physical mechanism responsible for generating ECME closely resembles the auroral emissions observed on planets \citep{2015SSRv..187...99B,2011JGRA..116.4212L} and on ultra-cool dwarfs \citep{2012ApJ...760...59N}.
The similarity extends to ultracool dwarfs and their exoplanets \citep{2019MNRAS.488..633V}. 
Furthermore, \citet{2021MNRAS.507.1979L} derived a scaling relationship of the non-thermal radio luminosity from Jupiter, through ultracool dwarfs, to early-type stars, which shows that the mechanism is similar in all these celestial bodies.
The resemblance of coherent radio emission to auroral emission led to its designation as Auroral Radio Emission \citep[ARE,][]{2020MNRAS.493.4657L}.
Stars on the main sequence exhibiting ECME radiation are alternatively designated as Main-Sequence Radio Pulse emitters (MRPs).
Currently, from literature 17 stars are classified as MRPs \citep{2022MNRAS.517.5756D,2022ApJ...925..125D}.

The magnetospheric processes in hot stars resemble auroral activity in Earth and giant planets \citep{2011ApJ...739L..10T}. Such magnetospheres are proverbial for their emission lines of molecules or neutral oxygen, which typically appear in UV region \citep{2014P&SS..103..291M,2016Icar..264..398S,2017Icar..284..264G}. Magnetic chemically peculiar stars show emission lines in optical spectra \citep{2004A&A...425..263C,2017PASJ...69...48S}, but they are considered to occur due to NLTE effects \citep{2016MNRAS.462.1123A,2020MNRAS.493.6095M}.
\citet{2019A&A...625A..34K} obtained dedicated phase-resolved HST spectroscopy of CU~Vir and searched for auroral lines. However, the search turned out to be negative for unclear reasons.
Therefore, the auroral emission lines have not yet been detected in magnetic hot stars.

There has never been a systematic search for these lines in hot stars.
To remedy the situation, we present a systematic study of auroral lines formation in the spectra of hot stars resulting from impacting electrons.

The modelling of the effect of impacting electrons is partly motivated by the X-ray radiation during solar flares.
Solar flares are known for their multi-thermal nature \citep{Nagasawa2022}.
In our study, the effect of impacting electrons is modelled by X-ray irradiation of the stellar atmosphere with blackbody radiation.

The paper is organized as follows. In Section~\ref{search}, we first search for auroral emission in FUSE data taken from the MAST archive.
Next in Section~\ref{sec:inputTLSY}, we present a description of the input parameters in our model atmospheres utilized for irradiation. 
Then in Section~\ref{physicalchange} we mentioned physical changes in the irradiated model atmosphere and compare it with analytical relation.
Subsequently, in Sections~\ref{sec:uv-spectrum}--~\ref{sec:emisIR}, we offer an overview of the emission lines which emerged from the irradiation in our models in UV, optical IR regions, and also show a list of emission lines which appear irrespective of irradiation in the hottest model considered.
Furthermore, in Section~\ref{emisssionlinehe}, we focus on the most prominent emission line observed in the highly irradiated spectra and establish the connection with population changes of ions in the irradiated models.
Based on this context, we derived the minimum required irradiation for the observation of the most prominent emission line and compared it with the kinetic energy of the wind taken from models and also with the energy emitted from magnetic stars in the radio region.

\section{Search for emission lines in FUSE spectra}
\label{search}

The unsuccessful search for auroral lines in HST spectra of CU Vir by \cite{2019A&A...625A..34K} motivated us to look for their presence in other similar objects.

We selected the FUSE satellite, which has a relatively high spectral resolution suitable for the search of narrow emission lines.
To obtain the list of magnetic stars with available far-UV spectra, we cross-referenced a list of magnetic stars from literature \citep[][]{2021yCat..36520031B,2019MNRAS.489.5669P} with a list of stars observed by FUSE satellite available in MAST archive.
However, only 6 stars met these criteria,  presented in Table \ref{tab:FUSE}.

Subsequently, we downloaded spectral data of all magnetic stars supplemented with spectra of stars with similar spectral types for reference (listed in Appendix in Table~\ref{tab:nonBstar}).
We compared the spectra of magnetic stars with three reference stars and with one non-irradiated spectrum with effective temperature which was closest to the temperature of a magnetic star.
For comparison we scaled spectra to unity within a range $1118-1120$\ \AA\,(Fig.~\ref{fig:HD200775}, the remaining spectra are given in Appendix~\ref{appendixFUSE}, Figs.~\ref{fig:HD47777}--\ref{fig:HD200311}).
Our focus was directed towards the identification of possible emission lines.
To mitigate the impact of complex continuum variation, the spectra were subdivided into smaller segments for manual meticulous inspection.

However, due to high noise levels and smaller exposure time than referenced stars, it is notably challenging to classify anything as an emission feature.
This challenge underpins our decision not to classify any feature as a potential emission.

\begin{figure}
	
    \includegraphics[width=\columnwidth]{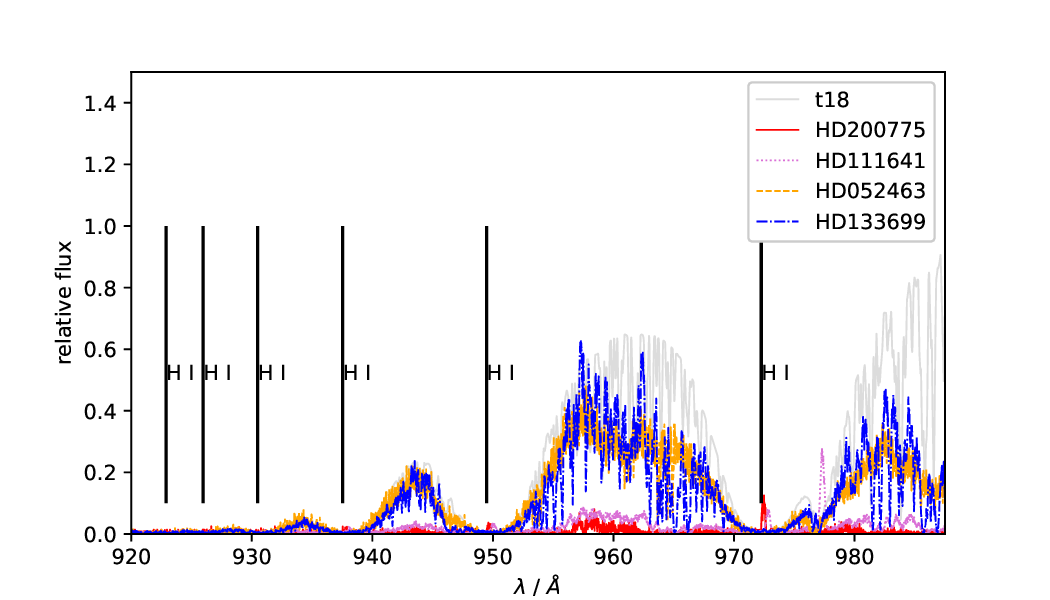}
    \includegraphics[width=\columnwidth]{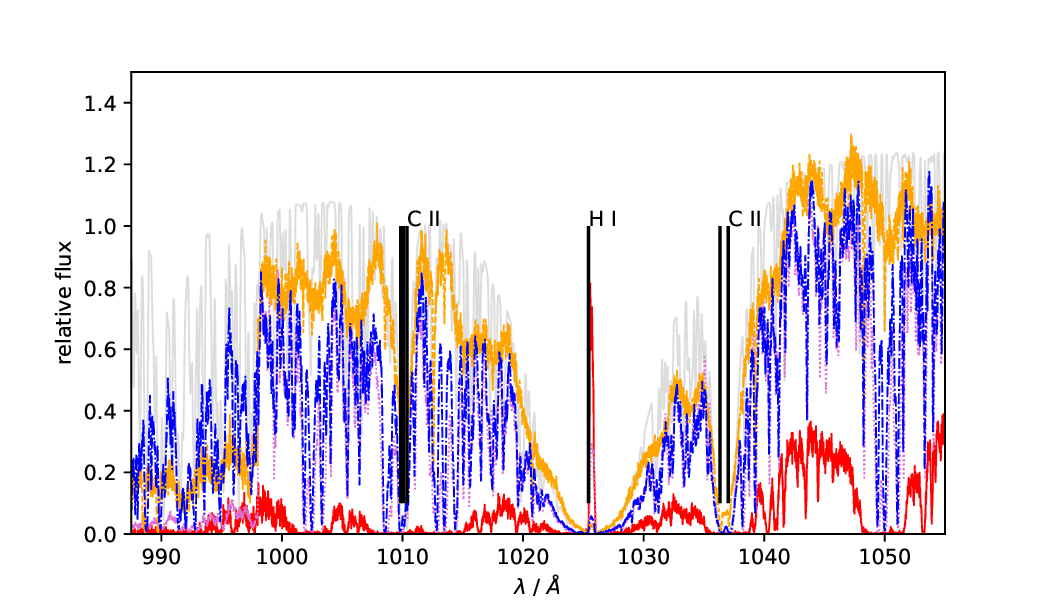}
    \includegraphics[width=\columnwidth]{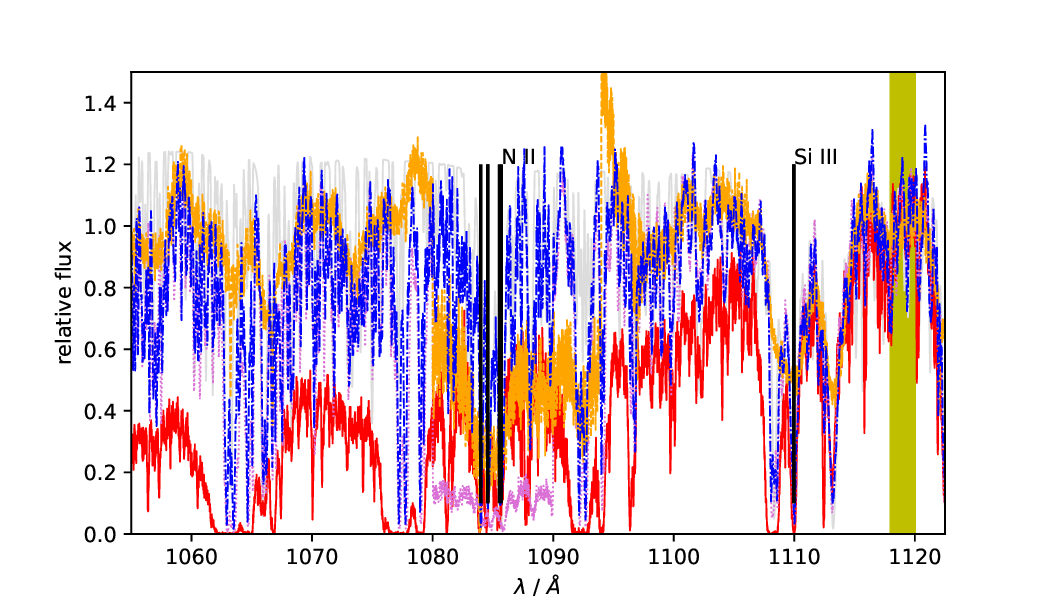}
    \includegraphics[width=\columnwidth]{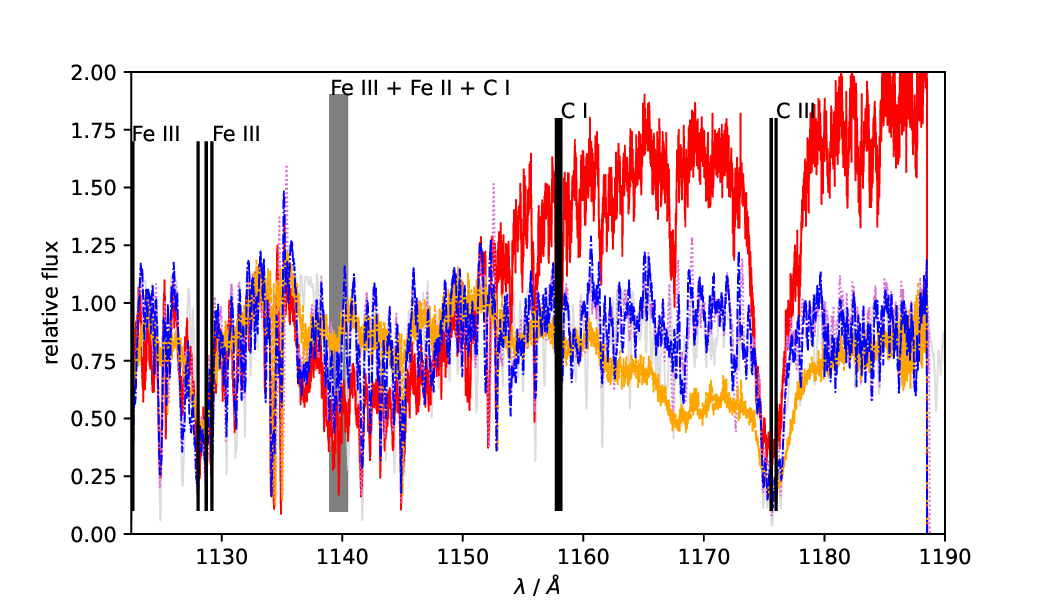}
    \caption{The normalized flux of the magnetic star HD 200775, alongside reference stars and the t18 model. The spectra of all observed stars have been rescaled to unity within the wavelength range of $1118-1120$\,\AA\ (indicated by yellow filled box), with respect to the t18 model. Prominent absorption lines have also been identified and annotated with black vertical lines.}
    \label{fig:HD200775}
\end{figure}





\begin{table}
    \centering
    \caption{List of magnetic stars found in FUSE catalogue.}
    
    \begin{tabular}{l|c|c|c}
    \hline
    ID & Sp. Type (Simbad) & Reference\\
    \hline
        HD 23478 & B3IV &  \cite{2021yCat..36520031B}\\
        HD 37151 & B8V & \cite{2021yCat..36520031B} \\
        HD 47777 & B3V & \cite{2021yCat..36520031B}\\
        HD 176386 & B9V & \cite{2021yCat..36520031B}\\
        HD 200311 & B9V & \cite{2021yCat..36520031B}\\
        HD 200775 & B2Ve (Herbig obj.) & \cite{2021yCat..36520031B}\\
        \hline
    \end{tabular}
    \label{tab:FUSE}
\end{table}

\section{Description of irradiated model atmospheres}
\label{simul}
Lack of success in searching for auroral emission in the previous section motivated us to calculate the theoretical models explaining the absence of X-ray emission.
We modelled the effect of fast electrons coming from magnetospheric reconnection by external X-ray irradiation.
Although the exact state of the atmosphere differs for cases of the impact of electrons and X-ray irradiation, the subsequent processes of recombination and deexcitation do not depend on the mechanism of ionization and excitation.

\subsection{Model atmospheres and synthetic spectra}
\label{sec:inputTLSY}

We calculated stellar model atmospheres from scratch using the code TLUSTY\footnote{TLUSTY v. 200} \citep{1995ApJ...439..875H}.
The code calculates NLTE plane-parallel model atmospheres in hydrostatic and radiative equilibrium. 
The models were calculated for different effective temperatures 15, 18, 21 and 30 kK and surface gravity $\log(g/ 1\,\text{cm}\,\text{s}^{-2})=4$ with different amount of external irradiation (Table \ref{tab:strength}).
Metal abundances were set to solar values from \citet{2009ARA&A..47..481A}.
We calculated models with the following elements in NLTE: H, He, C, N, O, Ne, Mg, Al, Si, S, and Fe.

\begin{table}
    \label{tab:modelgrid}
    \centering
    \caption{Parameters of the models with maximum adopted irradiation.}
    \begin{tabular}{c|c|c|c}
    \hline
        Model & $T_\text{eff}$ [kK] & $W$ &  $\log(\frac{ F_\text{irrad}}{F_\text{bol}} )$\\
         \hline\
         t15w12 &15& $10^{-12}$ & $-$3.771  \\
         t18w5\_12 &18& $5\times10^{-12}$ & $-$3.388\\
         t21w12 &21& $10^{-12}$ & $-$4.355\\
         t30w10 &30& $10^{-10}$ & $-$2.975\\
         \hline 
    \end{tabular}
    \label{tab:strength}
\end{table}

We employed irradiation in code TLUSTY using photons from the blackbody with the temperature set to $T_\text{irrad}=10^7\,$K, while we changed the dilution factor which affects the amount of flux impacting the stellar atmosphere.
The dilution factor $W$ of the irradiating flux is applied in the code in the equation $I_{\text{irrad}}=W \times B(T_\text{irrad})$.
Here $B(T_\text{irrad})$ is the Planck function at the temperature $T_\text{irrad}$.
The atmosphere is irradiated only between the minimum $\nu_\text{min}=1\times10^{12}$\,Hz and maximum $\nu_\text{max}=5.5\times10^{16}$\,Hz frequencies of the corresponding TLUSTY model.
Therefore, the irradiating flux $F_{\text{irrad}}$ is given by an integral of the Planck function between $\nu_\text{min}$ and $\nu_\text{max}$.

%

From model atmospheres calculated by TLUSTY, we simulated synthetic spectra using the SYNSPEC\footnote{SYNSPEC v. 49} code 
\citep{1995ApJ...439..875H}.
We have computed spectra for several irradiated and non-irradiated cases in the wavelength range of 900\,\r{A} to $10^5\,$\r{A}.
We used two different line lists in SYNSPEC, one for wavelength under 7500\,\AA\, which is included from SYNSPEC web page\footnote{http://tlusty.oca.eu/Synspec49/synspec.html} and the second one above 7500\,\AA with primarily IR lines for the rest taken from the VALD database \citep{1995A&AS..112..525P}.
The turbulent velocity has been set to $2\,\text{km}\,\text{s}^{-1}$. 
We created the spectrum in SYNSPEC taking TLUSTY models as input with only including elements which were solved in NLTE in TLUSTY.
This was done to prevent false emission from the heated upper atmosphere which was present when we first calculated the spectrum including elements which SYNSPEC calculated only in LTE.
Subsequently, we applied the code ROTIN on calculated spectra to perform rotational and instrumental convolution.
We assumed rotation of the star with $v_{\text{rot}}=30\,\text{km}\,\text{s}^{-1}$.


\subsection{Physical changes in the irradiated model atmospheres}
\label{physicalchange}
We investigated changes induced by strong irradiation in stellar model atmospheres.
For the case of the atmosphere of a grey accretion disk with external irradiation, \citet[Eq. 3.23] {1990ApJ...351..632H} introduced a penetration depth
\begin{equation}
    \label{eq:penetrationdepth}
    \tau_{\text{pen}} = \frac{4}{3}W \left( \frac{T_{\text{irrad}}}{T_{\text{eff}}} \right)^{4},
\end{equation}
which separates regions of a dominant and weak influence of irradiation.
In Eq.~\ref{eq:penetrationdepth} we employed the corrected $T_\text{irrad}$, which corresponds to a black body that emits the same flux as used to irradiate model atmospheres.
The effect of irradiation becomes negligible for optical depths higher than the penetration depth, while the atmosphere is strongly heated by irradiation above the penetration depth.

For our specific cases, the penetration depth from Eq.~\eqref{eq:penetrationdepth} does not exceed 0.02, even for the hottest model with the strongest irradiation.
\begin{figure}
    \includegraphics[width=\columnwidth]{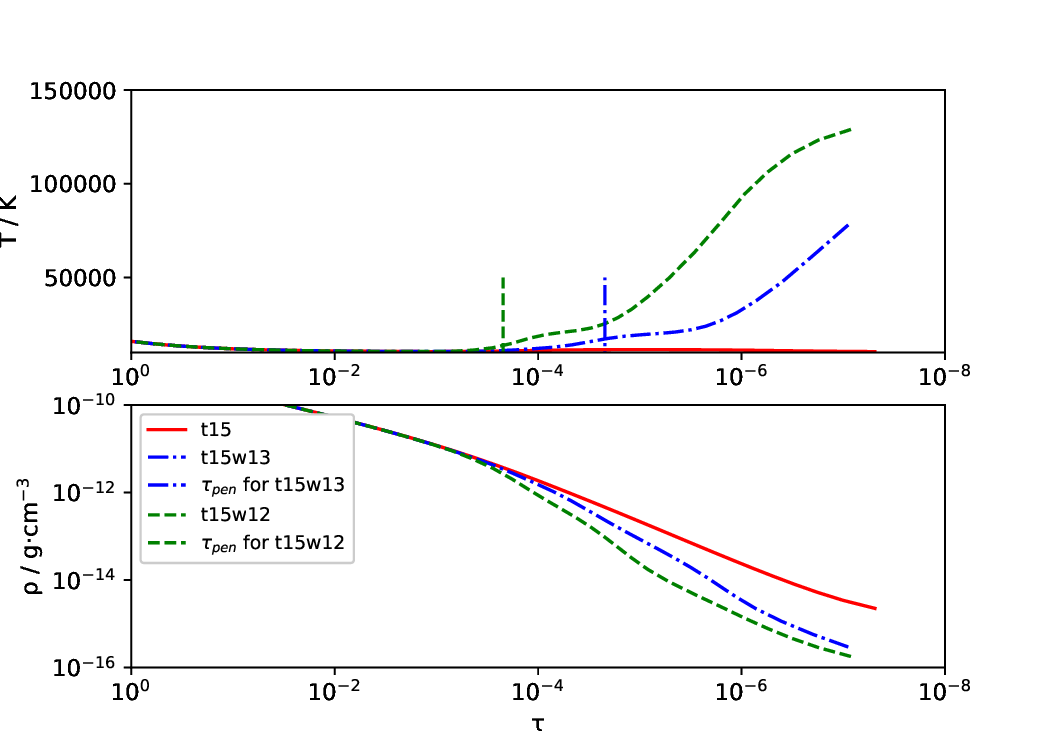}
    \caption{Variations of density and temperature as a function of the Rosseland mean optical depth in the models with and without irradiation. Plotted for the effective temperature 15\,kK. The red solid line denotes the model without irradiation, the green dashed line denotes irradiation with a dilution factor set to $1 \times 10^{-13}$, and the blue dashed-dotted line denotes irradiation with dilution factor set to $1 \times 10^{-12}$. 
    {\it Upper panel}: Variations of temperature. Vertical dashed/dashed-dotted lines represent the calculated value of penetration depth from Eq.\ref{eq:penetrationdepth} for specified dilution and flux $F_{\text{irrad}}$ from the irradiated body.
    {\it Lower panel}: Variations of density.
    }
    \label{fig:TD}
\end{figure}
This is demonstrated in Fig.~\ref{fig:TD}, where we compare the variations of temperature in irradiated and non-irradiated model atmospheres, and we also marked the penetration depth for specific models.
Only the outermost part of the atmosphere, where the Rosseland mean optical depth is significantly lower than one, experiences significant heating.
The optical depth of the region where the irradiation starts to heat the atmosphere nicely agrees with penetration depth from Eq.\ref{eq:penetrationdepth}.
Additionally, as a result of hydrostatic equilibrium, the heating process results in a slight decrease in density within irradiated regions.
Altered conditions have important implications for the modelling of stellar wind in irradiated stars due to adjustments of velocity and density at the base of the wind.

\begin{figure}
    \includegraphics[width=\columnwidth]{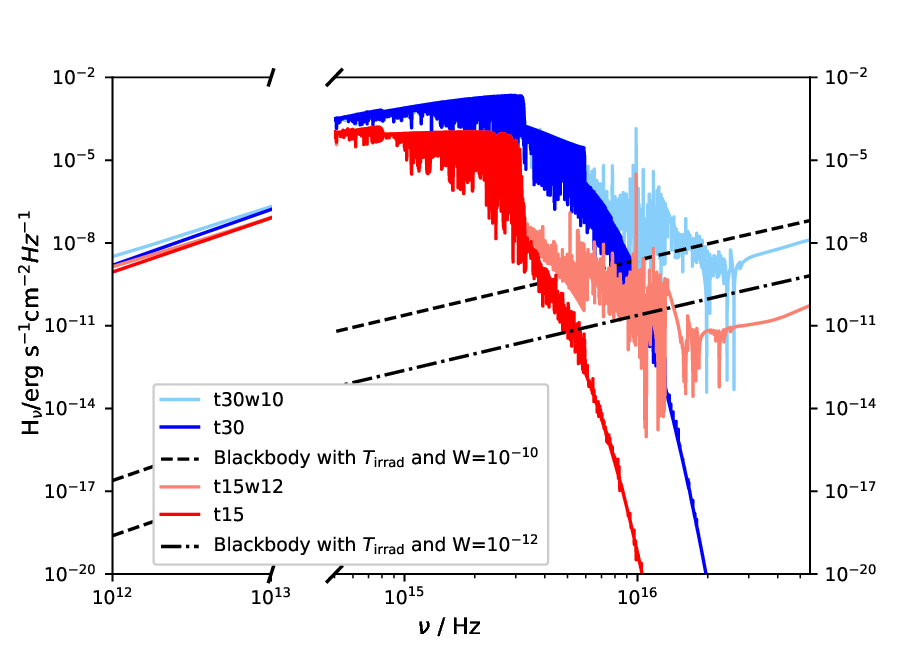}
    \caption{Emergent Eddington flux as a function of frequency for model atmospheres with and without irradiation. Dashed and dot-dashed lines correspond to the blackbody spectrum multiplied by a corresponding dilution factor.}
    \label{fig:bb_irrad}
\end{figure}

In Fig. \ref{fig:bb_irrad} we show flux energy distribution for t15 and t30 models. From the plots, it follows that the changes caused by irradiation appear primarily at high frequencies.

\begin{figure*}

    \includegraphics[width=\textwidth]{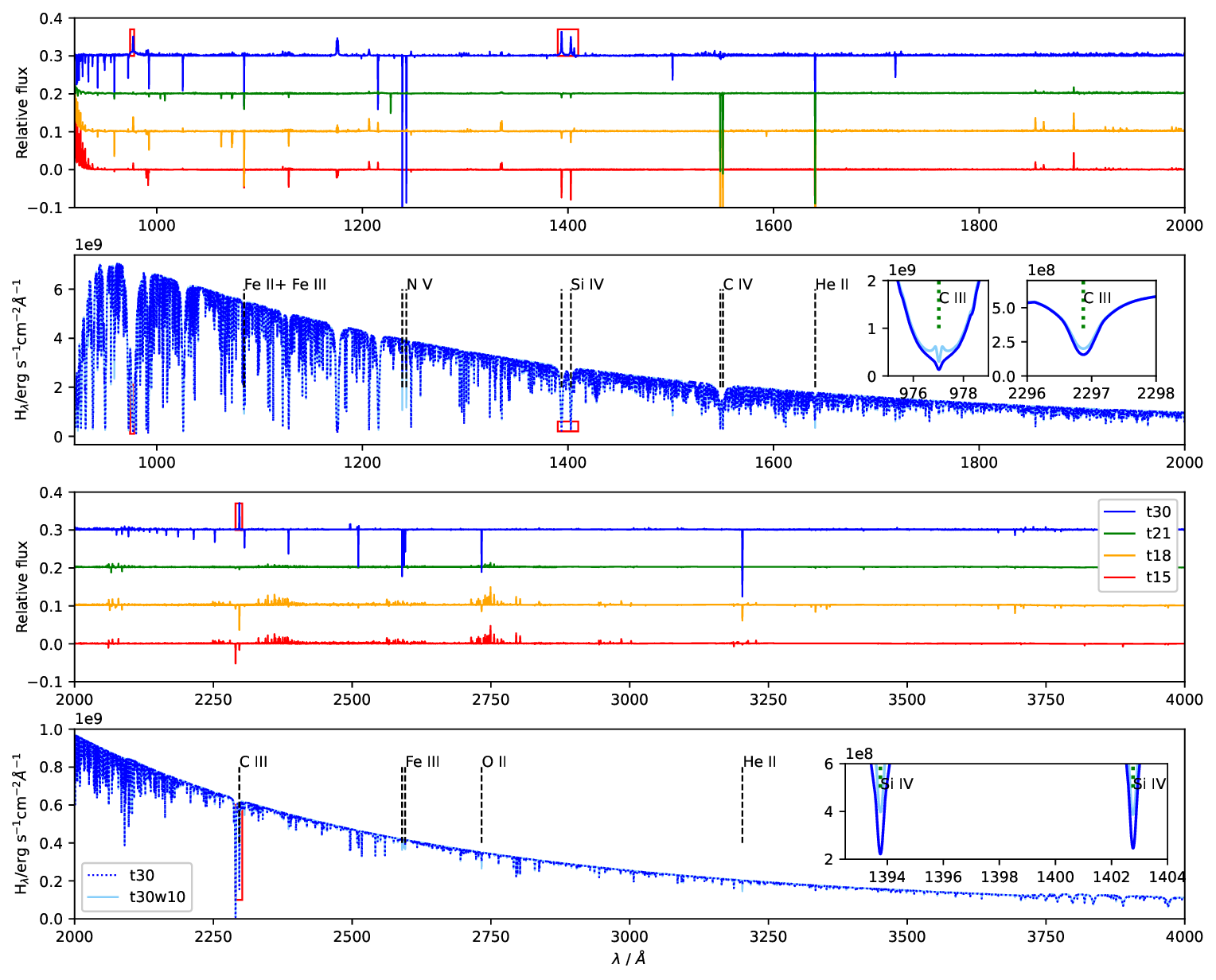}
    \caption{Comparison of irradiated and non-irradiated fluxes in the UV domain. The first and third panels give relative flux differences defined as the difference between irradiated model flux with adopted $W$ given in Table \ref{tab:strength} and non-irradiated flux divided by continuum. The plots for different effective temperatures are vertically shifted for better visibility. The second and fourth panels plot the Eddington fluxes of the t30 irradiated model (solid sky blue line) and non-irradiated model (blue dashed line).
    The black dashed vertical lines identify prominent lines.
    The insets show three zoomed-in parts of the Eddington flux in the regions where the relative flux shows at least a five per cent increase. They are also shown in the relative and also in absolute figures with red rectangles around them.}
    \label{fig:Optical+UV}
\end{figure*}

\begin{figure*}
    \includegraphics[width=\textwidth]{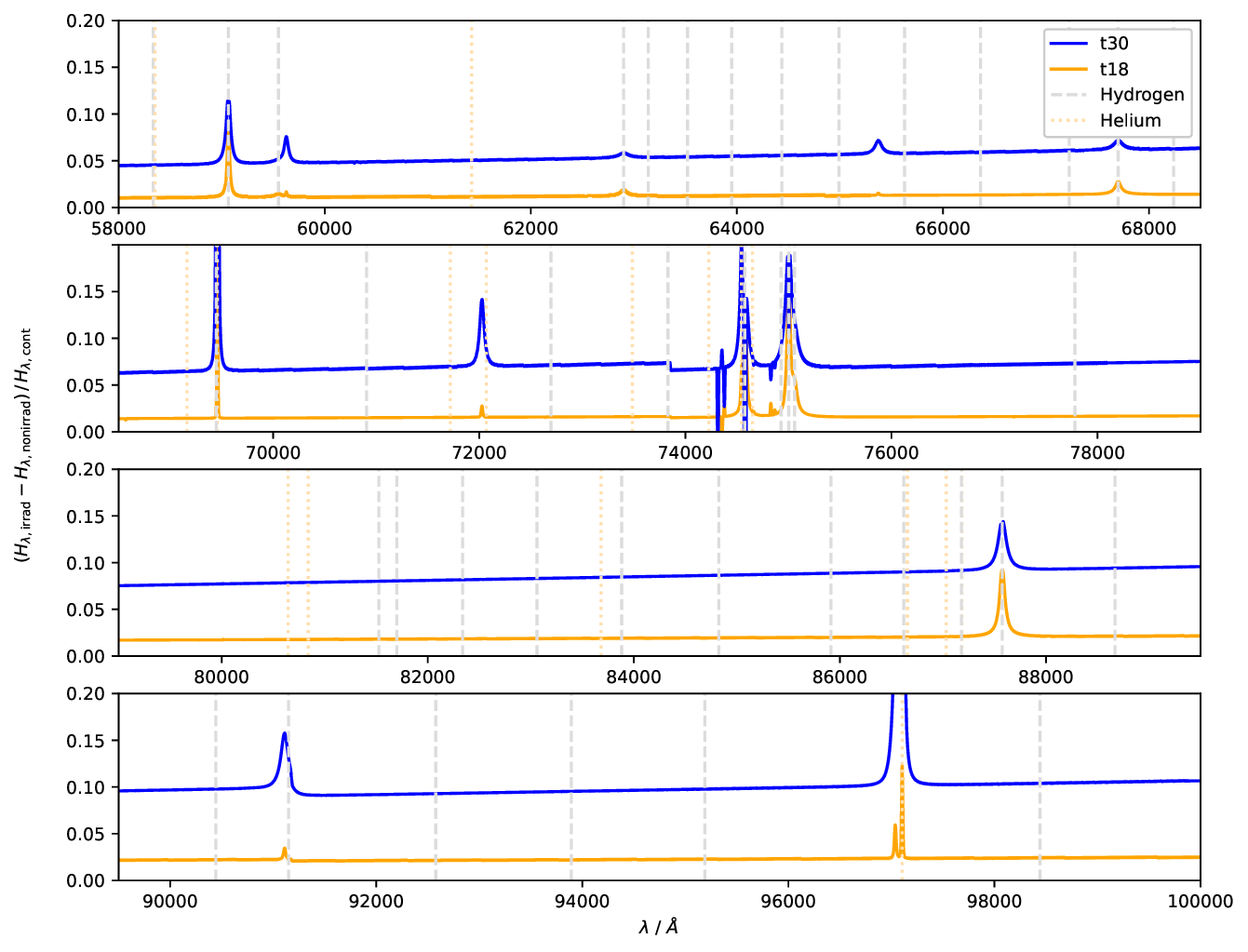}
    \caption{Relative flux differences defined as the difference between irradiated model flux with adopted $W$ given in Table \ref{tab:strength} and non-irradiated flux divided by continuum for t30 and t18 models. Selected hydrogen lines (grey dashed vertical lines) and helium lines (orange dotted vertical lines) are marked.
    The flux excess of the t30 model is more than five per cent of the theoretical continuum and increases with wavelength.} 
    \label{fig:compt18andt30}
\end{figure*}



\section{Search for emission lines in synthetic spectra}
\label{sec:emission}

\subsection{Change of the spectrum in UV}
\label{sec:uv-spectrum}

To search for emission lines in synthetic spectra, we calculated the difference in Eddington fluxes between the maximally irradiated and non-irradiated models from the SYNSPEC code.
This difference was then divided by the theoretical non-irradiated continuum, $(H_{\lambda \text{, irrad}} - H_{\lambda\text{, nonirrad}})/ H_{\lambda \text{, cont}}$, and plotted as a function of wavelength in Figs.~\ref{fig:Optical+UV} and \ref{fig:compt18andt30}.
To focus on significant emission lines, we adopted an arbitrary threshold of five per cent for the selection of emission lines from the relative spectra. 
When selecting this value we took into account challenges encountered in practical spectrum analysis related to factors such as wind, reddening, and so on.
We also understand that the value of the threshold together with rotational convolution can increase or decrease the number of emission lines found, as we observed that higher rotational convolution stretches the emission.

Initially, our focus was directed towards the UV part of the spectrum.
A comparison between emergent fluxes in the UV domain is shown in Fig.~\ref{fig:Optical+UV}.
We identified a few lines with emission features in the cores of absorption lines, nevertheless, their intensity is lower than the hypothetical continuum at a given wavelength. 
The most prominent of these emission features are shown in insets in Fig.~\ref{fig:Optical+UV}.
Several irradiated models exhibited lines with significantly stronger absorption features, particulary prominent were a doublet \ion{N}{v} $\lambda\,1242$, numerous blended \ion{Fe}{ii} and \ion{Fe}{iii} lines near $\lambda\,1084$, a doublet \ion{Si}{iv} $\lambda\,1402$, a doublet \ion{C}{iv} $\lambda\,1550$, \ion{He}{ii} $\lambda\,1640$, \ion{C}{iii} $\lambda\,2296$, \ion{O}{ii} $\lambda\,2733$, and \ion{He}{ii} $\lambda\,3203$.
However, in the case of \ion{Si}{iv} $\lambda\,1402$ for the irradiated t30 model, there was the inverse effect and the absorption feature was weaker.
This suggests that irradiation has the potential to induce stronger absorption in a few elements.
Even without consideration of X-ray irradiation, the identification of most of the emission or absorption features in UV stellar spectra poses a significant challenge, primarily attributable to the interplay of numerous factors.
These include the determination of abundance, effective temperature and gravity, the presence of weak stellar wind, instrumental noise, and the influence of NLTE effects.
Precise determination at a fine spectral scale is hindered by these factors.
Moreover, the stronger or weaker absorption caused by irradiated spectra introduces another layer of complexity, making it even more susceptible to misinterpretation.
To summarize this effort, we did not find any significant emission lines in the UV spectral region in our set of irradiated models.
This explains the missing UV emission lines in CU Vir \citep{2019A&A...625A..34K}.

\subsection{Lines in optical and NIR regions, and emission lines resulting from increased effective temperature}
\label{ss:onirem}
We found no significant emission line created by irradiation, stronger than five per cent, compared to non-irradiated models within the 2500--18000 \AA\ wavelength range.

Still, in the t30 model, we identified a few emissions that appeared either as absorption lines or as a continuum without features in cooler models.
These lines were found even in the model without irradiation, therefore, they are not caused by irradiation as per se but rather by the increase of effective temperature, indicating NLTE effects.
Emission lines stronger by more than five percent than continuum only in t30 model even without irradiation include \ion{C}{ii} $\lambda 9903$~\AA, \ion{He}{I} $\lambda 10830$~\AA, \ion{Si}{ii} $\lambda 13395$~\AA, \ion{Si}{iii} $\lambda 13644$~\AA, \ion{H}{I} $\lambda 18750$~\AA, \ion{He}{i} $\lambda 21655$~\AA, \ion{He}{i} $\lambda 26252$~\AA.
Additionally, in models under irradiation, the majority of these lines weakened with increasing irradiation.

In the case of the H$\alpha$ $\lambda 6562$~\AA \, line, the X-ray irradiation leads to a stronger absorption in the t18 and t30 models.
For the t30 model, additional absorption of \ion{He}{II} $\lambda 6559$ \AA \, becomes visible in the wing of H$\alpha$.

\subsection{Emission lines in long-wavelength infrared }
\label{sec:emisIR}
In the extended search to the infrared (IR) part of the spectrum from $18000$ to $10^{5}$ \AA, we found many candidate emission lines.
All potential candidate lines are listed in Table  \ref{tab:candidate} sorted based on the number of models where the lines are present.
Because models t18w5\_12 and t30w10 are the most irradiated, we plotted them in Fig.~\ref{fig:compt18andt30}, where we show relative Eddington flux plotted in the IR part of the spectrum. 
Except for the emission lines, we observed a relative increase in the continuum flux with a wavelength in both models.
The corresponding IR excess is especially prominent in the case of the t30w10 model because the relative flux difference is higher than the adopted threshold of five per cent for wavelengths longer than $6\,\mu$m.

\begin{table}
    \centering
    \caption{Candidates emission lines from models.}
    \begin{tabular}{c|c|c}
    \hline
        Present in model/s & Candidate emission line wavelength in \AA\\
         \hline
         t15, t18, t21, t30 & 69458 (\ion{He}{ii})\\
         t15, t18, t30 &75003 (\ion{H}{i}), 97104 (\ion{He}{ii})\\
         t15, t30 & 40493 (\ion{He}{II})\\
         t18, t30 & 37394 (\ion{H}{i}), 46524 (\ion{H}{i}), 55810 (\ion{He}{ii}),\\
         & 59064 (\ion{H}{i}), 74576 (\ion{H}{i}),\\
          &  74585 (\ion{H}{i}),   87575 (\ion{H}{i})\\
         t30 UV & 976 (\ion{C}{iii}), 1393 (\ion{Si}{iv}), 1402 (\ion{Si}{iv}), \\
         & 2296 (\ion{C}{iii}) \\
         t30 IR & 18742, 28251, 30945,\\
         (All \ion{He}{ii})& 40510, 42171, 60960+$^a$ \\
         \hline
         

\multicolumn{3}{l}{$^a$ Whole continuum for irradiated t30 model is higher than threshold}\\
\multicolumn{3}{l}{for these wavelengths.}\\
    \end{tabular}

    \label{tab:candidate}

\end{table}

\begin{figure}
    \includegraphics[width=\columnwidth]{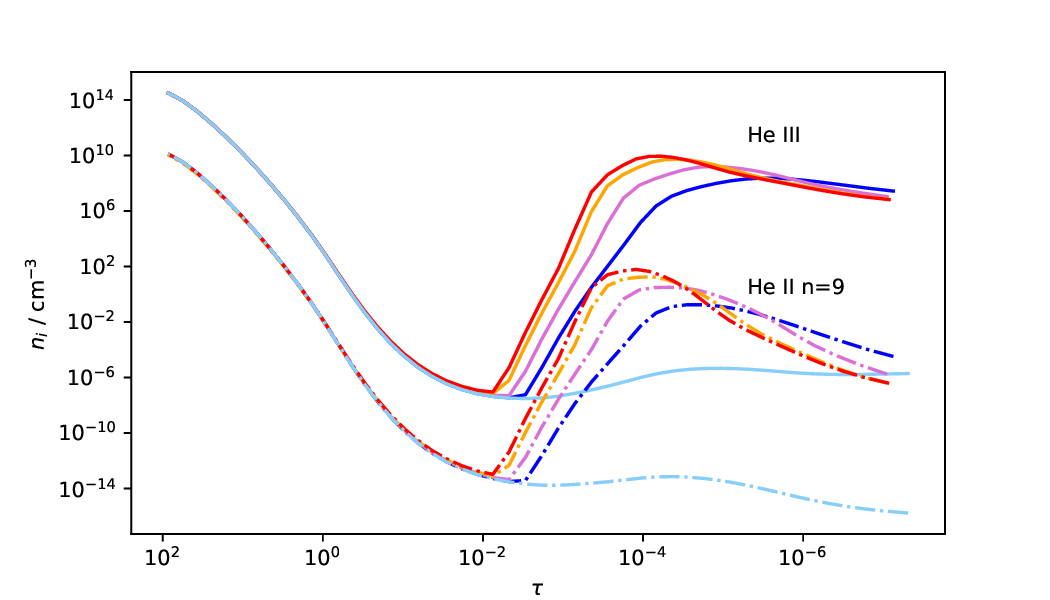}
    \includegraphics[width=\columnwidth]{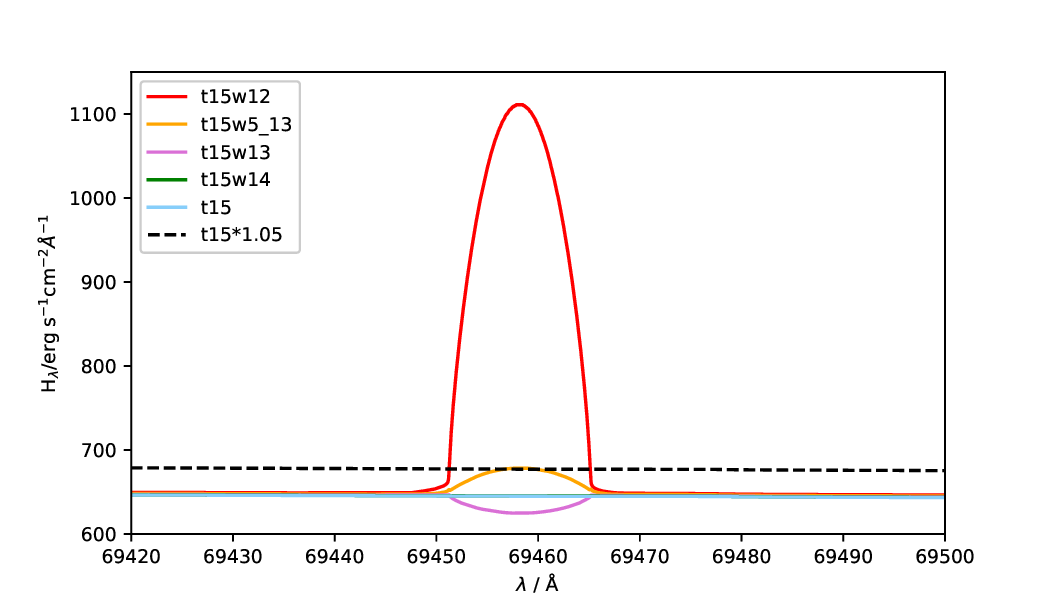} 
    \caption{Model t15. {\it Upper panel}: Population of \ion{He}{iii} (solid lines) and \ion{He}{ii} (dash-dotted lines) with a principal quantum number equal to nine for different values of irradiation as a function of Rosseland optical depth. Same colour as the bottom panel. The stronger irradiation causes higher ionisation. The label w13 denotes W=$10^{-13}$ and w5\_13 denotes W=$5\times10^{-13}$ etc. {\it Bottom panel}: Eddington flux around \ion{He}{ii} 69458 \AA\,line for different values of irradiation. Dashed line marks adopted a threshold for classifying emissions as observable.}
    \label{fig:cand15}
\end{figure}

\subsection{The most prominent line}
\label{emisssionlinehe}
For a more detailed study, we focused on the most prominent candidate for the emission line.
We selected the \ion{He}{ii} $69458$ \AA\ emission line which was found in all irradiated models.
This line corresponds to the transition between levels with principal quantum numbers 8 and 9.
In comparison, we can also find strong emission at $97104$ \AA\, in models t18 and t30, which represents the transition line of \ion{He}{ii} between $n=9$ and $n=10$.
Nevertheless, the transition line of \ion{He}{ii} generated from $n=7$ to $n=8$ is at $47620$ {\AA}
%
was not found to generate emissions in any model.

\begin{figure}
    \includegraphics[width=\columnwidth]{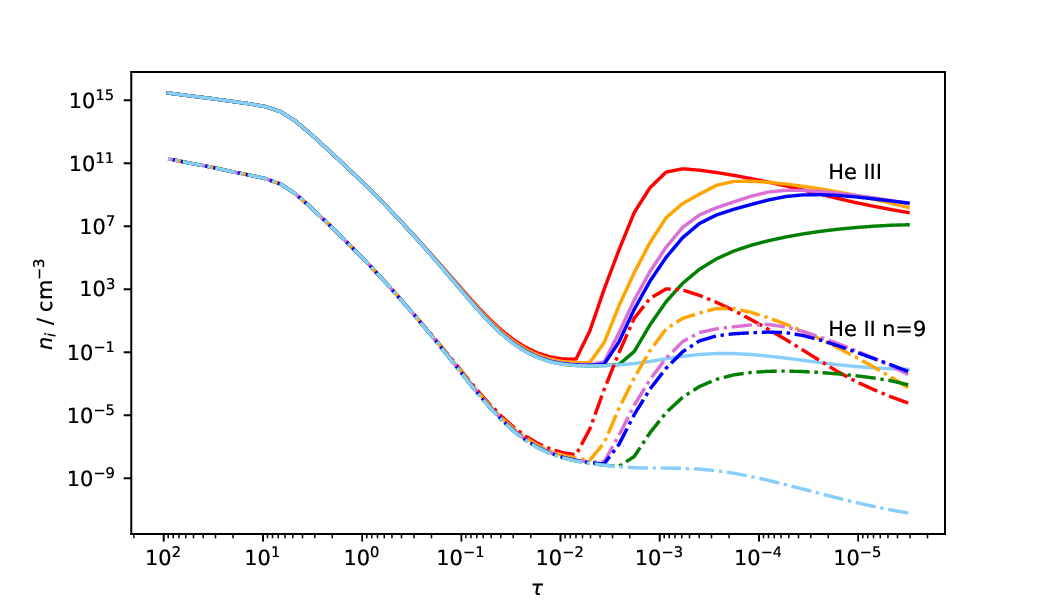}
    \includegraphics[width=\columnwidth]{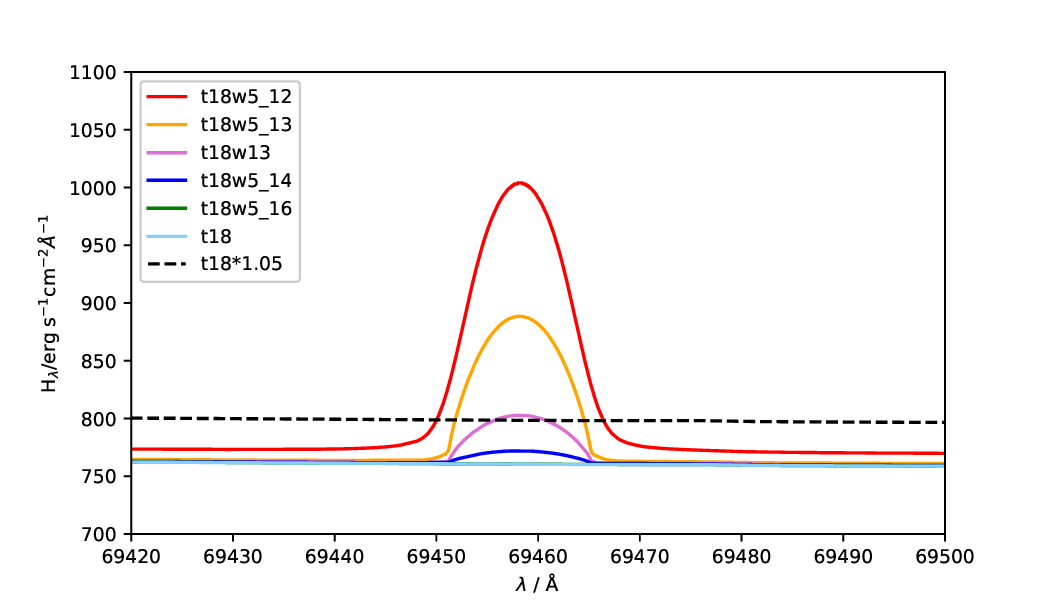} 
    \caption{Same as Figure \ref{fig:cand15} but for t18 model.}
    \label{fig:cand18}
\end{figure}

\begin{figure}

    \includegraphics[width=\columnwidth]{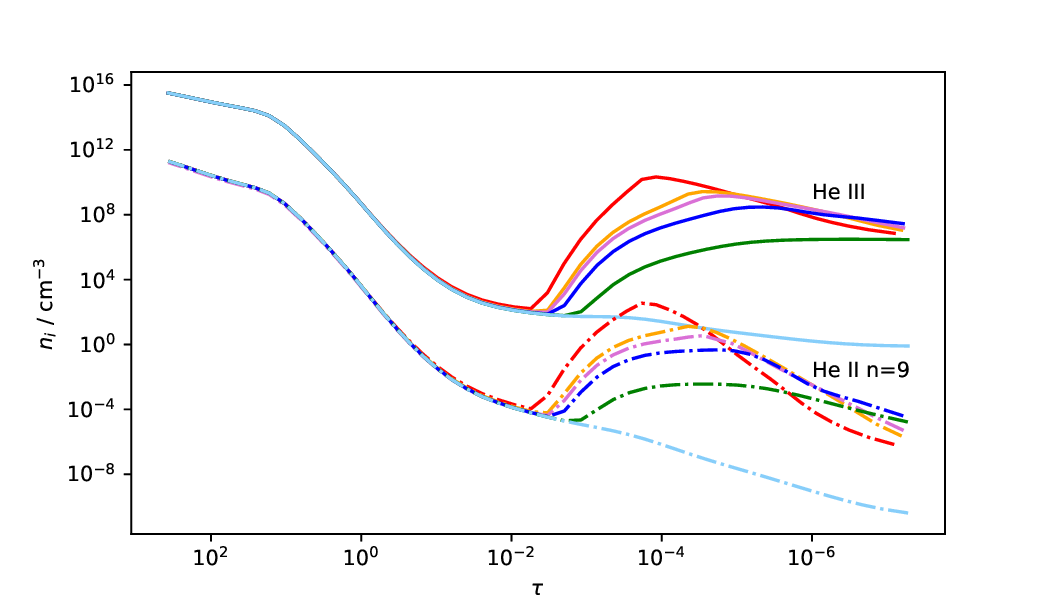}
    \includegraphics[width=\columnwidth]{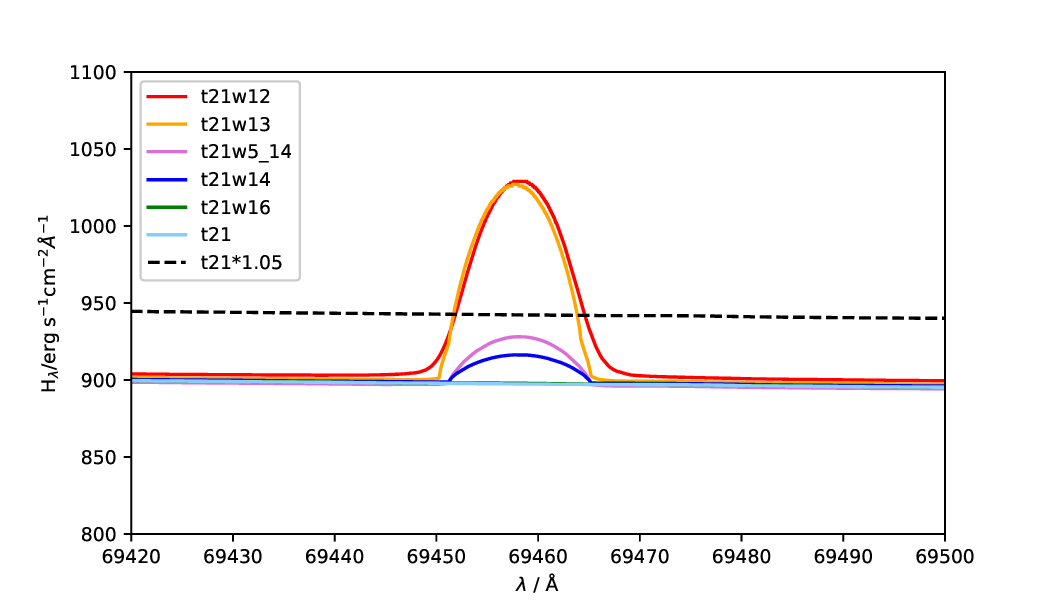} 
    \caption{Same as Figure \ref{fig:cand18} but for t21 model.}
    \label{fig:cand21}
\end{figure}

\begin{figure}

    \includegraphics[width=\columnwidth]{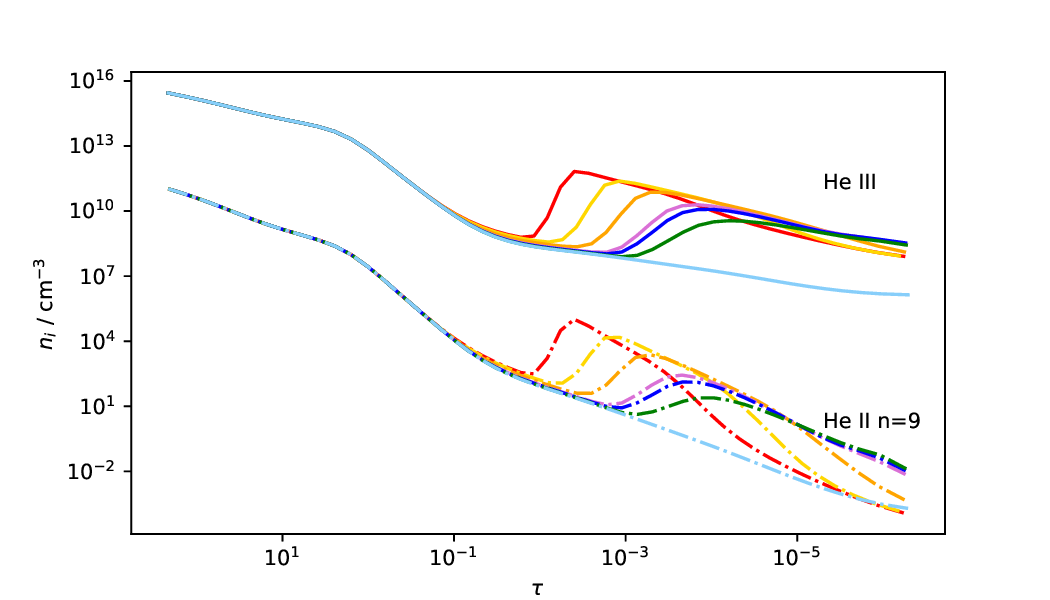}
    \includegraphics[width=\columnwidth]{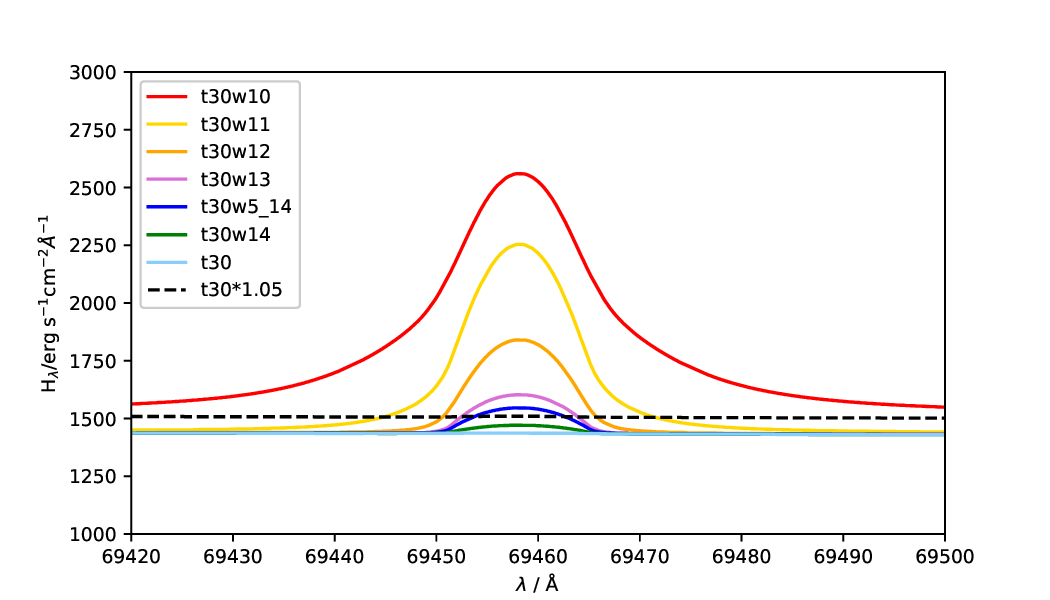} 
    \caption{Same as Figure \ref{fig:cand18} but for t30 model.}
    \label{fig:cand30}
\end{figure}

We inspected changes in the population of \ion{He}{ii} in $n=9$ level and \ion{He}{iii}, and at the same time, we checked the Eddington flux around \ion{He}{ii} $69458$\ \AA\ line.
This is shown for models t18, t21, and t30 in Figs.~\ref{fig:cand15} -- \ref{fig:cand30}.
From the figures, it follows that the upper layers of irradiated atmospheres are not just heated but the population of these states is by ten orders of magnitude higher than in non-irradiated cases.

We also inspected b-factors of various levels of \ion{He}{ii}.
In colder non-irradiated models b-factors tend to hover around unity even within regions characterized by lower Rosseland optical depth $\tau<10^{-4}$.
This trend also explains the adequateness of  LTE models for colder stars.
In contrast, for the hottest model in the non-irradiated case, the b-factors exhibit a more complex, but descending pattern for all ionization levels, and only in the case of $n=20$ level, which is the highest level considered in TLUSTY, the b-factor is near unity.
However, in the case of irradiated models, the b-factors for \ion{He}{II} are significantly higher reaching up to $10^6$ for $n=1$ within the low-optical depth region.
As the excitation energy increases, the b-factors significantly decrease.
Notably for $n>5$, the b-factors consistently remain below $\sim2$.
Levels with higher excitation energies tend to have more complex behaviour around $\tau \sim 10^{-4}$, marking the region where the temperature starts to increase in comparison with non-irradiated models.
But as the opacity decreases, collective behaviour becomes evident, they are approaching unity similarly as if this region was in LTE.

\begin{figure}

    \includegraphics[width=\columnwidth]{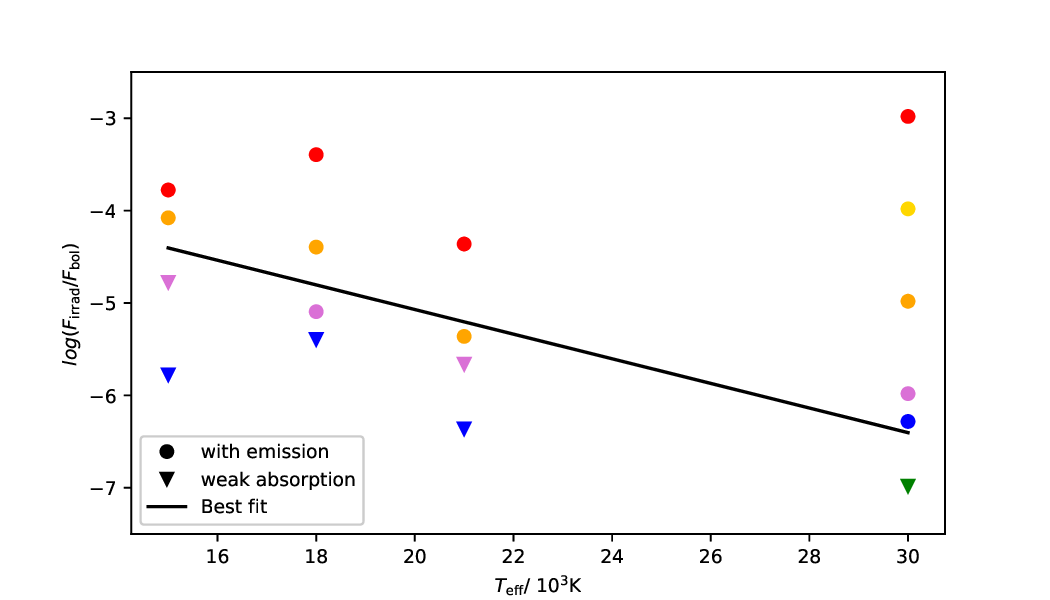}
    \includegraphics[width=\columnwidth]{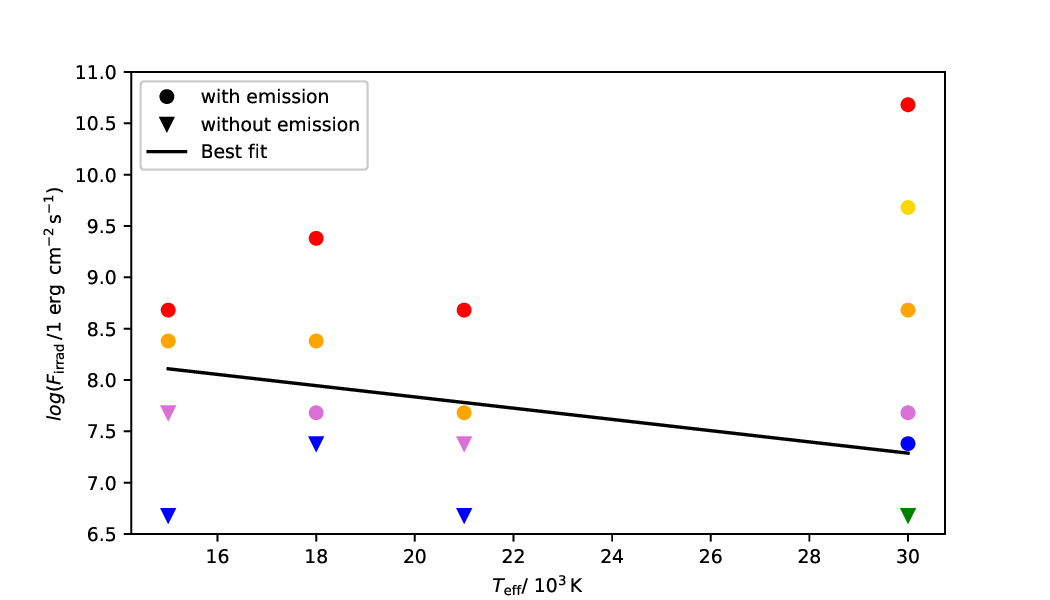}
    \caption{Grid of calculated models. We plot X-ray irradiation as a function of effective temperature. Models with emission lines present in the spectra are marked using circles with the same colour as in Fig.\,\ref{fig:cand18}-\ref{fig:cand30}. Models where the emission is absent are marked using triangles. The black solid line is the fit Eqs.~\eqref{eqfxbol} and \eqref{eqfxerg} of the minimum irradiation where the emission emission is present. The fit is based on t18, t21, and t30 models. {\it Upper panel}: Ordinate is the logarithm of the ratio of irradiated flux to the flux of the model. {\it Bottom panel}: Ordinate is the logarithm of the irradiated flux.}
    \label{fig:compALL}
\end{figure}

We used the transition \ion{He}{ii} $69458$ \AA\, to calculate the minimum required irradiation for which this line can be observed.
We fitted minimum irradiation for which was emission shown versus the effective temperature of models.
The fit is
\begin{equation}
\label{eqfxbol}
    \log( F_{\text{irrad}} / F_{\text{bol}}) =- 2.406-0.1333
\times (T_{\text{eff}}/10^3\, \text{K}), 
\end{equation}
where $F_{\text{bol}}$ is bolometric flux of model with given temperature and $F_\text{irrad}$ is irradiated flux.
For a given temperature, we can calculate the ratio of irradiated flux to bolometric flux required to observe the emission line.
This relationship can be expressed as a linear fit of irradiation as a function of the effective temperature of the models, yielding the equation 
\begin{equation}
\label{eqfxerg}
    \log(F_{\text{irrad}}/ 1\,\text{erg}\,\text{cm}^{-2}\,\text{s}^{-1}) =8.931-0.05479\times(T_{\text{eff}}/10^3\, \text{K})\,.
\end{equation}
We also plotted this fit in Fig.~\ref{fig:compALL} for better representation.
From this fit, we conclude that the required irradiation for generating an emission line is lower for a higher effective temperature of the star.

\begin{figure}
    \includegraphics[width=\columnwidth]{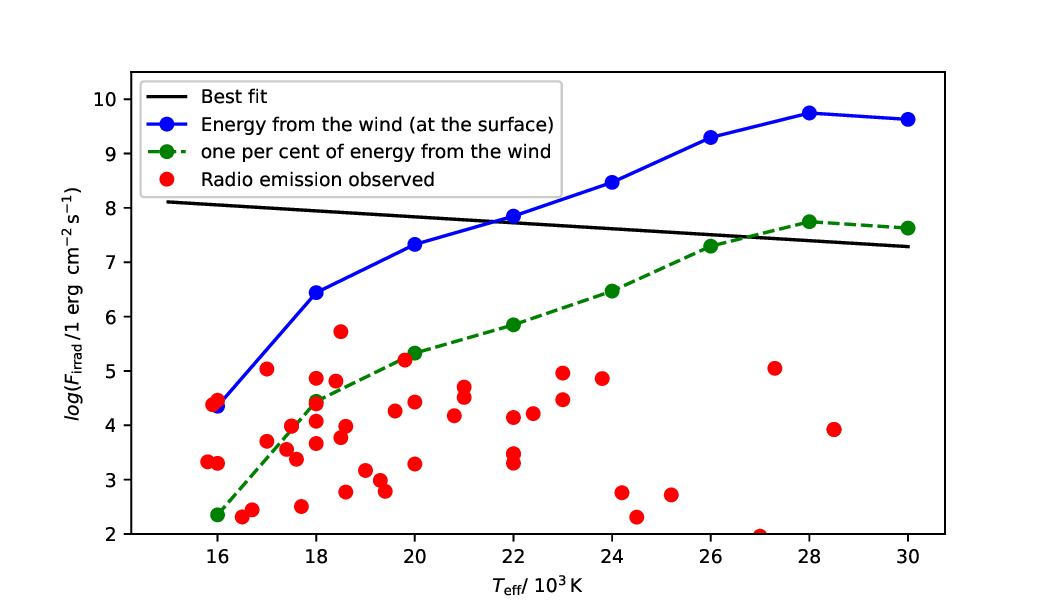}
    \caption{Comparison of the fit (solid black line) of the irradiation required to observe emission lines with energy generated in the wind \citep[blue solid line with models marked with a circle,][]{ 2014psce.conf..250K}.
    Red circles are integrated radio emissions from observations collected by \citet{2022MNRAS.513.1429S}.
    The green dashed line represents one per cent of energy from the wind.}
    \label{fig:winds}
\end{figure}

We also compared the minimum flux required to generate emission with stellar wind kinetic energy flux $\dot{M} v_{\infty}^2 / (8 \pi R^2) $ predicted for solar-metallicity main-sequence B stars \citep{2014psce.conf..250K} and  mean radio flux observed in magnetic early-type stars \citep{2022MNRAS.513.1429S} 
in Fig.~\ref{fig:winds}.
To calculate the radio flux, we derived the radius of the star from the luminosity and effective temperature given in \citet{2022MNRAS.513.1429S}.

Based on our models (Fig.~\ref{fig:winds}), we concluded that the energy delivered from the wind is capable of inducing emission in stars with effective temperature exceeding $22$\,kK.
However, if, for instance, only one per cent of the wind energy can be converted into irradiation energy, then emission can be observed in stars with effective temperatures higher than $27$\,kK.

These findings may provide insight into understanding the challenges encountered in uncovering auroral lines, as demonstrated by  \citet{2019A&A...625A..34K}.
In their study of CU Vir ($T_{\text{eff}}\approx13$\,kK), attempts to identify auroral lines in the UV region of CU Vir proved unsuccessful.
Our models indicate that for this star, characterized by low effective temperature and weak observed X-ray emission \citep{2018A&A...619A..33R}, the X-ray intensity alone is not sufficient to generate emission lines.
Instead of emission lines, the irradiation in the UV part of the spectrum would manifest by stronger absorption features in a few specific lines.
From this, it follows that hotter single or binary stars are more suitable candidates to search for emission lines, because the energy converted from mass loss in hotter stars could be sufficient to generate observable auroral lines.

\section{Discussion}

Our finding suggests that the emission lines due to intense X-ray irradiation may appear mainly in the IR domain.
This is in agreement with results in literature \citep[][eq. 5.12]{2014tsa..book.....H}, which state that within the IR regime, where the energy difference between two levels $l$ and $u$ divided by Boltzmann constant times the temperature is much less than one, or in physics notation $h\nu_{lu} /kT \ll 1$, emission in lines stemming from NLTE effects becomes notably more probable.
From our model, the irradiation influences the formation of emission lines in the IR domain, in particular of hydrogen and helium.

Existing literature has reported the occurrence of X-ray emission in B-type stars with luminosities on the order of $\log({L_{\text{X}}}/L_{\text{bol}})=-7$ \citep{2018A&A...619A..33R}.
Notably, there are also stars with even stronger X-ray emission with $\log({L_{\text{X}}}/L_{\text{bol}})>-5$ \citep{1996A&AS..118..481B}. 
Based on our work, these stars have the potential to exhibit infrared emission lines as effects of X-ray irradiation.
However, we want to point out that our computational models revealed more subtle effects, the weaker and stronger absorption features in different lines in the UV portion of the spectrum, which can probably also be detectable.

%

A critical assumption underlying our work is that the effect of impacting electrons on the atmosphere can be adequately approximated with a weakened blackbody irradiation with a specific temperature.
But for example in the Sun, the bremsstrahlung radiation is produced when the plasma particles, which are accelerated in the magnetic reconnection, inject the solar chromosphere.
In this region two types of bremsstrahlung are produced: hard X-rays (above 20keV) by relativistic particles and soft X-rays by strongly heated but thermal plasma \citep{1993SoPh..146..177D}.
However, during the solar flare and activity, the coronal plasma is strongly heated and is expected to emit thermal radiation.
Similar consideration can be also relevant for stellar activity \citep{udDoula2006}.
However, in our model, we assumed that the particles released from the reconnection do not significantly penetrate the stellar atmosphere and their energy is reemitted as electromagnetic radiation.
While this can be true in the solar case, relativistic electrons may have different penetration depths and different spectra, which is bremsstrahlung in nature.
This is different from photons, whose penetration depth is given by opacity, and they are not repelled by electric charge.
In the future, we plan to test the effect of modification of irradiation spectra and analyse relativistic impacting electrons.
In any case, the energy transferred from impacting electrons and photons should be equal.

Irradiation in TLUSTY was also modelled by \citet{2016A&A...586A.146V}. 
Their irradiating flux had an effective temperature of $42$\,kK and dilution factor was much higher than in our case (in the order of one hundredth). Therefore, our model has much more energetic but more diluted irradiation.
Despite these differences, we both found that the physical changes in temperature, density and pressure deep inside the star were negligible in comparison with non-irradiated models.
Moreover, despite different irradiation parameters, the irradiation was changing the spectrum very subtly (Németh, private communication).

A similar analysis of irradiated hydrogen NLTE model atmospheres of B stars was conducted in \citet{2000ASPC..214..705K}, but with only hydrogen.
Irradiation in both analyses affected the outer layers of models, but, in our case, the same irradiation caused the same increase in the temperature versus mass in the outer parts in all models independent of the effective temperature of the model.
We attribute the differences to the fact that we use models including elements heavier than hydrogen.
As opposed to the case of the pure hydrogen model, in models with heavy elements, most atoms are not fully ionized in t30 models, and, consequently, may still show emission lines.
Additionally, hydrogen lines in proximity to $7.5\mu$m\, did not exhibit a monotonic correlation with increasing irradiation across all effective temperatures.
Particularly, in the case of the irradiated t30 model, we observed a more complex relationship with irradiation.

\citet{2000A&A...363.1055M} modelled a spectrum of a B3 V star irradiated by a thermal X-ray source with $T_{\text{irrad}}=10^{8}$\,K. 
However, they focused on the UV and X-ray part of the spectra including only hydrogen, helium, and iron in LTE.
Because of that the comparison of spectra cannot be done.
Comparison of temperature versus optical depth showed a very similar pattern to our results, that is, heating of the outer parts of the atmosphere.
In their case with higher irradiation $\log W \sim -16$ hydrogen behaved as in thermal equilibrium and caused the disappearance of the Lyman jump.
However, they assumed a significantly higher irradiation flux than we included in our analysis, because we account just for a fraction $q$ of irradiating black body given by the maximum frequency of the models.
Our models did not converge for that high irradiation. For instance, the t18 model would require $\log W \sim -16-4 \log {\left( \frac{T_{\text{rad}}}{T_{\text{irrad}}} \right)}-\log q \sim -9$) to observe the disappearance of Lyman jump.
 After analysis of the flux immediately below $912$ \AA, we identified that the flux increased in the irradiated versus non-irradiated model in all cases.
For the cooler models, the increase was most visible and was approximately three times for the t15 model, two times in the t18 model, and roughly twenty per cent for the t21 model. 
For the t30 model, flux increased only in the order of per cent.

\section{Conclusions}
In this study, we searched for potential emission features in the FUSE spectra of magnetic stars, which contain spectral regions near the Lyman limit, and also analysed irradiated synthetic spectra.
Our goal was to identify any potential auroral emission lines in spectra.
In FUSE spectra we did not classify any emission features that could be attributed to the auroral emission of magnetic stars.

We conducted a comprehensive analysis and search for auroral lines in the synthetic spectra of OB stars.
These auroral lines are a direct consequence of the ionization of the atmosphere resulting from the impacting electrons generated during magnetospheric reconnection events.
To examine the impact of these electrons, we replaced electrons with a simplified X-ray irradiation model.

We found out that model atmospheres with X-ray irradiation show a significant increase in the temperature of the upper layers of the stellar atmospheres.
This caused only a subtle effect in the UV region, specifically weak emission or absorption components appearing in the centres of a few strong absorption lines.

The effects of high-energy irradiation are more pronounced in the long-wavelength region and lead to the appearance of several emission lines and infrared excess. 
In the IR part of the spectra, we compiled a list of potential emission lines.
We selected the most prominent line \ion{He}{ii} 69458 \AA, which was observed in all our irradiated models and used it for the next analysis.
Based on this prominent \ion{He}{ii} 69458 \AA\ line, we determined the minimum irradiation threshold necessary to observe the emission. 

Subsequently, we calculated the best fit for the minimal required irradiation as a function of effective temperature, based on the basic assumption that the required irradiation solely depends on the effective temperature of the model.
Fit shows that the required irradiation for observing the most prominent line decreases with increasing effective temperature, meaning that for hotter stars less irradiation is needed for the appearance of emission.
Admittedly, hot stars can exhibit very strong radiatively driven winds, which can also serve as a source of emission in line and obstruct the detection of the auroral lines.

\section*{Acknowledgements}
We would like to thank Dr.~Filip Hroch for the maintenance of the old MIRSAM server, which was used for the main part of the calculation of models, and for restoring the server after one of us had accidentally halted it. 
Next, we want to thank Dr.~Jan Benáček for the beneficial communication about Solar flares.
We also thank Dr.~Péter Németh for discussing the influence of irradiation on stars and for sharing his experience with irradiation in code TLUSTY.
We thank Dr.~Ján Budaj and Dr.~Martin Piecka for many stimulating conversations.
And last but not least, we thank Dr.~Ivan Hubený for his time and his willingness to help with difficult questions about code TLUSTY.

\section*{Data Availability}
TLUSTY and SYNSPEC are open-source software.
The spectroscopic data underlying this work are free to download from the MAST archive and model atmospheres generated from TLUSTY will be shared on reasonable request to the corresponding author.




\bibliographystyle{mnras}
\bibliography{example} 

\begin{thebibliography}{}
\makeatletter
\relax
\def\mn@urlcharsother{\let\do\@makeother \do\$\do\&\do\#\do\^\do\_\do\%\do\~}
\def\mn@doi{\begingroup\mn@urlcharsother \@ifnextchar [ {\mn@doi@}
  {\mn@doi@[]}}
\def\mn@doi@[#1]#2{\def\@tempa{#1}\ifx\@tempa\@empty \href
  {http://dx.doi.org/#2} {doi:#2}\else \href {http://dx.doi.org/#2} {#1}\fi
  \endgroup}
\def\mn@eprint#1#2{\mn@eprint@#1:#2::\@nil}
\def\mn@eprint@arXiv#1{\href {http://arxiv.org/abs/#1} {{\tt arXiv:#1}}}
\def\mn@eprint@dblp#1{\href {http://dblp.uni-trier.de/rec/bibtex/#1.xml}
  {dblp:#1}}
\def\mn@eprint@#1:#2:#3:#4\@nil{\def\@tempa {#1}\def\@tempb {#2}\def\@tempc
  {#3}\ifx \@tempc \@empty \let \@tempc \@tempb \let \@tempb \@tempa \fi \ifx
  \@tempb \@empty \def\@tempb {arXiv}\fi \@ifundefined
  {mn@eprint@\@tempb}{\@tempb:\@tempc}{\expandafter \expandafter \csname
  mn@eprint@\@tempb\endcsname \expandafter{\@tempc}}}

\bibitem[\protect\citeauthoryear{{Alexeeva}, {Ryabchikova}  \&
  {Mashonkina}}{{Alexeeva} et~al.}{2016}]{2016MNRAS.462.1123A}
{Alexeeva} S.~A.,  {Ryabchikova} T.~A.,   {Mashonkina} L.~I.,  2016, \mn@doi
  [\mnras] {10.1093/mnras/stw1635}, \href
  {https://ui.adsabs.harvard.edu/abs/2016MNRAS.462.1123A} {462, 1123}

\bibitem[\protect\citeauthoryear{{Asplund}, {Grevesse}, {Sauval}  \&
  {Scott}}{{Asplund} et~al.}{2009}]{2009ARA&A..47..481A}
{Asplund} M.,  {Grevesse} N.,  {Sauval} A.~J.,   {Scott} P.,  2009, \mn@doi
  [\araa] {10.1146/annurev.astro.46.060407.145222}, \href
  {https://ui.adsabs.harvard.edu/abs/2009ARA&A..47..481A} {47, 481}

\bibitem[\protect\citeauthoryear{{Badman}, {Branduardi-Raymont}, {Galand},
  {Hess}, {Krupp}, {Lamy}, {Melin}  \& {Tao}}{{Badman}
  et~al.}{2015}]{2015SSRv..187...99B}
{Badman} S.~V.,  {Branduardi-Raymont} G.,  {Galand} M.,  {Hess} S. L.~G.,
  {Krupp} N.,  {Lamy} L.,  {Melin} H.,   {Tao} C.,  2015, \mn@doi [\ssr]
  {10.1007/s11214-014-0042-x}, \href
  {https://ui.adsabs.harvard.edu/abs/2015SSRv..187...99B} {187, 99}

\bibitem[\protect\citeauthoryear{{Berghoefer}, {Schmitt}  \&
  {Cassinelli}}{{Berghoefer} et~al.}{1996}]{1996A&AS..118..481B}
{Berghoefer} T.~W.,  {Schmitt} J.~H.~M.~M.,   {Cassinelli} J.~P.,  1996, \aaps,
  \href {https://ui.adsabs.harvard.edu/abs/1996A&AS..118..481B} {118, 481}

\bibitem[\protect\citeauthoryear{{Bychkov}, {Bychkova}  \& {Madej}}{{Bychkov}
  et~al.}{2021}]{2021yCat..36520031B}
{Bychkov} V.~D.,  {Bychkova} L.~V.,   {Madej} J.,  2021, VizieR Online Data
  Catalog, \href {https://ui.adsabs.harvard.edu/abs/2021yCat..36520031B} {pp
  J/A+A/652/A31}

\bibitem[\protect\citeauthoryear{{Castelli} \& {Hubrig}}{{Castelli} \&
  {Hubrig}}{2004}]{2004A&A...425..263C}
{Castelli} F.,  {Hubrig} S.,  2004, \mn@doi [\aap]
  {10.1051/0004-6361:20041011}, \href
  {https://ui.adsabs.harvard.edu/abs/2004A&A...425..263C} {425, 263}

\bibitem[\protect\citeauthoryear{{Das}, {Chandra}, {Shultz}, {Leto},
  {Mikul{\'a}{\v{s}}ek}, {Petit}  \& {Wade}}{{Das}
  et~al.}{2022a}]{2022MNRAS.517.5756D}
{Das} B.,  {Chandra} P.,  {Shultz} M.~E.,  {Leto} P.,  {Mikul{\'a}{\v{s}}ek}
  Z.,  {Petit} V.,   {Wade} G.~A.,  2022a, \mn@doi [\mnras]
  {10.1093/mnras/stac3123}, \href
  {https://ui.adsabs.harvard.edu/abs/2022MNRAS.517.5756D} {517, 5756}

\bibitem[\protect\citeauthoryear{{Das} et~al.,}{{Das}
  et~al.}{2022b}]{2022ApJ...925..125D}
{Das} B.,  et~al., 2022b, \mn@doi [\apj] {10.3847/1538-4357/ac2576}, \href
  {https://ui.adsabs.harvard.edu/abs/2022ApJ...925..125D} {925, 125}

\bibitem[\protect\citeauthoryear{{Dennis} \& {Zarro}}{{Dennis} \&
  {Zarro}}{1993}]{1993SoPh..146..177D}
{Dennis} B.~R.,  {Zarro} D.~M.,  1993, \mn@doi [\solphys] {10.1007/BF00662178},
  \href {https://ui.adsabs.harvard.edu/abs/1993SoPh..146..177D} {146, 177}

\bibitem[\protect\citeauthoryear{{Gustin}, {Grodent}, {Radioti}, {Pryor},
  {Lamy}  \& {Ajello}}{{Gustin} et~al.}{2017}]{2017Icar..284..264G}
{Gustin} J.,  {Grodent} D.,  {Radioti} A.,  {Pryor} W.,  {Lamy} L.,   {Ajello}
  J.,  2017, \mn@doi [\icarus] {10.1016/j.icarus.2016.11.017}, \href
  {https://ui.adsabs.harvard.edu/abs/2017Icar..284..264G} {284, 264}

\bibitem[\protect\citeauthoryear{{Hubeny}}{{Hubeny}}{1990}]{1990ApJ...351..632H}
{Hubeny} I.,  1990, \mn@doi [\apj] {10.1086/168501}, \href
  {https://ui.adsabs.harvard.edu/abs/1990ApJ...351..632H} {351, 632}

\bibitem[\protect\citeauthoryear{{Hubeny} \& {Lanz}}{{Hubeny} \&
  {Lanz}}{1995}]{1995ApJ...439..875H}
{Hubeny} I.,  {Lanz} T.,  1995, \mn@doi [\apj] {10.1086/175226}, \href
  {https://ui.adsabs.harvard.edu/abs/1995ApJ...439..875H} {439, 875}

\bibitem[\protect\citeauthoryear{{Hubeny} \& {Mihalas}}{{Hubeny} \&
  {Mihalas}}{2014}]{2014tsa..book.....H}
{Hubeny} I.,  {Mihalas} D.,  2014, {Theory of Stellar Atmospheres}.
Princeton Univ. Press

\bibitem[\protect\citeauthoryear{{Krti{\v{c}}ka}}{{Krti{\v{c}}ka}}{2014}]{2014psce.conf..250K}
{Krti{\v{c}}ka} J.,  2014, in {Mathys} G.,  {Griffin} E.~R.,  {Kochukhov} O.,
  {Monier} R.,   {Wahlgren} G.~M.,  eds, Putting A Stars into Context:
  Evolution, Environment, and Related Stars. pp 250--255

\bibitem[\protect\citeauthoryear{{Krti{\v{c}}ka} et~al.,}{{Krti{\v{c}}ka}
  et~al.}{2019}]{2019A&A...625A..34K}
{Krti{\v{c}}ka} J.,  et~al., 2019, \mn@doi [\aap]
  {10.1051/0004-6361/201834937}, \href
  {https://ui.adsabs.harvard.edu/abs/2019A&A...625A..34K} {625, A34}

\bibitem[\protect\citeauthoryear{{Kub{\'a}t}}{{Kub{\'a}t}}{2000}]{2000ASPC..214..705K}
{Kub{\'a}t} J.,  2000, in {Smith} M.~A.,  {Henrichs} H.~F.,   {Fabregat} J.,
  eds,  Astronomical Society of the Pacific Conference Series Vol. 214, IAU
  Colloq. 175: The Be Phenomenon in Early-Type Stars. p.~705

\bibitem[\protect\citeauthoryear{{Lamy} et~al.,}{{Lamy}
  et~al.}{2011}]{2011JGRA..116.4212L}
{Lamy} L.,  et~al., 2011, \mn@doi [Journal of Geophysical Research (Space
  Physics)] {10.1029/2010JA016195}, \href
  {https://ui.adsabs.harvard.edu/abs/2011JGRA..116.4212L} {116, A04212}

\bibitem[\protect\citeauthoryear{{Leto} et~al.,}{{Leto}
  et~al.}{2020}]{2020MNRAS.493.4657L}
{Leto} P.,  et~al., 2020, \mn@doi [\mnras] {10.1093/mnras/staa587}, \href
  {https://ui.adsabs.harvard.edu/abs/2020MNRAS.493.4657L} {493, 4657}

\bibitem[\protect\citeauthoryear{{Leto} et~al.,}{{Leto}
  et~al.}{2021}]{2021MNRAS.507.1979L}
{Leto} P.,  et~al., 2021, \mn@doi [\mnras] {10.1093/mnras/stab2168}, \href
  {https://ui.adsabs.harvard.edu/abs/2021MNRAS.507.1979L} {507, 1979}

\bibitem[\protect\citeauthoryear{{Madej} \& {R{\'o}za{\'n}ska}}{{Madej} \&
  {R{\'o}za{\'n}ska}}{2000}]{2000A&A...363.1055M}
{Madej} J.,  {R{\'o}za{\'n}ska} A.,  2000, \aap, \href
  {https://ui.adsabs.harvard.edu/abs/2000A&A...363.1055M} {363, 1055}

\bibitem[\protect\citeauthoryear{{Marcolino}, {Bouret}, {Sundqvist}, {Walborn},
  {Fullerton}, {Howarth}, {Wade}  \& {ud-Doula}}{{Marcolino}
  et~al.}{2013}]{2013MNRAS.431.2253M}
{Marcolino} W.~L.~F.,  {Bouret} J.~C.,  {Sundqvist} J.~O.,  {Walborn} N.~R.,
  {Fullerton} A.~W.,  {Howarth} I.~D.,  {Wade} G.~A.,   {ud-Doula} A.,  2013,
  \mn@doi [\mnras] {10.1093/mnras/stt323}, \href
  {https://ui.adsabs.harvard.edu/abs/2013MNRAS.431.2253M} {431, 2253}

\bibitem[\protect\citeauthoryear{{Mashonkina}}{{Mashonkina}}{2020}]{2020MNRAS.493.6095M}
{Mashonkina} L.,  2020, \mn@doi [\mnras] {10.1093/mnras/staa653}, \href
  {https://ui.adsabs.harvard.edu/abs/2020MNRAS.493.6095M} {493, 6095}

\bibitem[\protect\citeauthoryear{{Molyneux} et~al.,}{{Molyneux}
  et~al.}{2014}]{2014P&SS..103..291M}
{Molyneux} P.~M.,  et~al., 2014, \mn@doi [\planss] {10.1016/j.pss.2014.08.007},
  \href {https://ui.adsabs.harvard.edu/abs/2014P&SS..103..291M} {103, 291}

\bibitem[\protect\citeauthoryear{{Morel} et~al.,}{{Morel}
  et~al.}{2015}]{2015IAUS..307..342M}
{Morel} T.,  et~al., 2015, in {Meynet} G.,  {Georgy} C.,  {Groh} J.,   {Stee}
  P.,  eds,  IAU Symp. Vol. 307, New Windows on Massive Stars. pp 342--347
  (\mn@eprint {arXiv} {1408.2100}), \mn@doi{10.1017/S1743921314007054}

\bibitem[\protect\citeauthoryear{{Nagasawa}, {Kawate}, {Narukage}, {Takahashi},
  {Caspi}  \& {Woods}}{{Nagasawa} et~al.}{2022}]{Nagasawa2022}
{Nagasawa} S.,  {Kawate} T.,  {Narukage} N.,  {Takahashi} T.,  {Caspi} A.,
  {Woods} T.~N.,  2022, \mn@doi [\apj] {10.3847/1538-4357/ac7532}, \href
  {https://ui.adsabs.harvard.edu/abs/2022ApJ...933..173N} {933, 173}

\bibitem[\protect\citeauthoryear{{Naz{\'e}}, {Sundqvist}, {Fullerton},
  {ud-Doula}, {Wade}, {Rauw}  \& {Walborn}}{{Naz{\'e}}
  et~al.}{2015}]{2015MNRAS.452.2641N}
{Naz{\'e}} Y.,  {Sundqvist} J.~O.,  {Fullerton} A.~W.,  {ud-Doula} A.,  {Wade}
  G.~A.,  {Rauw} G.,   {Walborn} N.~R.,  2015, \mn@doi [\mnras]
  {10.1093/mnras/stv1445}, \href
  {https://ui.adsabs.harvard.edu/abs/2015MNRAS.452.2641N} {452, 2641}

\bibitem[\protect\citeauthoryear{{Nichols}, {Burleigh}, {Casewell}, {Cowley},
  {Wynn}, {Clarke}  \& {West}}{{Nichols} et~al.}{2012}]{2012ApJ...760...59N}
{Nichols} J.~D.,  {Burleigh} M.~R.,  {Casewell} S.~L.,  {Cowley} S.~W.~H.,
  {Wynn} G.~A.,  {Clarke} J.~T.,   {West} A.~A.,  2012, \mn@doi [\apj]
  {10.1088/0004-637X/760/1/59}, \href
  {https://ui.adsabs.harvard.edu/abs/2012ApJ...760...59N} {760, 59}

\bibitem[\protect\citeauthoryear{{Oksala}, {Grunhut}, {Kraus}, {Borges
  Fernandes}, {Neiner}, {Condori}, {Campagnolo}  \& {Souza}}{{Oksala}
  et~al.}{2015}]{2015A&A...578A.112O}
{Oksala} M.~E.,  {Grunhut} J.~H.,  {Kraus} M.,  {Borges Fernandes} M.,
  {Neiner} C.,  {Condori} C.~A.~H.,  {Campagnolo} J.~C.~N.,   {Souza} T.~B.,
  2015, \mn@doi [\aap] {10.1051/0004-6361/201525987}, \href
  {https://ui.adsabs.harvard.edu/abs/2015A&A...578A.112O} {578, A112}

\bibitem[\protect\citeauthoryear{{Owocki}, {Shultz}, {ud-Doula}, {Sundqvist},
  {Townsend}  \& {Cranmer}}{{Owocki} et~al.}{2020}]{2020MNRAS.499.5366O}
{Owocki} S.~P.,  {Shultz} M.~E.,  {ud-Doula} A.,  {Sundqvist} J.~O.,
  {Townsend} R. H.~D.,   {Cranmer} S.~R.,  2020, \mn@doi [\mnras]
  {10.1093/mnras/staa2325}, \href
  {https://ui.adsabs.harvard.edu/abs/2020MNRAS.499.5366O} {499, 5366}

\bibitem[\protect\citeauthoryear{{Owocki}, {Shultz}, {ud-Doula}, {Chandra},
  {Das}  \& {Leto}}{{Owocki} et~al.}{2022}]{2022MNRAS.513.1449O}
{Owocki} S.~P.,  {Shultz} M.~E.,  {ud-Doula} A.,  {Chandra} P.,  {Das} B.,
  {Leto} P.,  2022, \mn@doi [\mnras] {10.1093/mnras/stac341}, \href
  {https://ui.adsabs.harvard.edu/abs/2022MNRAS.513.1449O} {513, 1449}

\bibitem[\protect\citeauthoryear{{Petit} et~al.,}{{Petit}
  et~al.}{2019}]{2019MNRAS.489.5669P}
{Petit} V.,  et~al., 2019, \mn@doi [\mnras] {10.1093/mnras/stz2469}, \href
  {https://ui.adsabs.harvard.edu/abs/2019MNRAS.489.5669P} {489, 5669}

\bibitem[\protect\citeauthoryear{{Piskunov}, {Kupka}, {Ryabchikova}, {Weiss}
  \& {Jeffery}}{{Piskunov} et~al.}{1995}]{1995A&AS..112..525P}
{Piskunov} N.~E.,  {Kupka} F.,  {Ryabchikova} T.~A.,  {Weiss} W.~W.,
  {Jeffery} C.~S.,  1995, \aaps, \href
  {https://ui.adsabs.harvard.edu/abs/1995A&AS..112..525P} {112, 525}

\bibitem[\protect\citeauthoryear{{Robrade}, {Oskinova}, {Schmitt}, {Leto}  \&
  {Trigilio}}{{Robrade} et~al.}{2018}]{2018A&A...619A..33R}
{Robrade} J.,  {Oskinova} L.~M.,  {Schmitt} J.~H.~M.~M.,  {Leto} P.,
  {Trigilio} C.,  2018, \mn@doi [\aap] {10.1051/0004-6361/201833492}, \href
  {https://ui.adsabs.harvard.edu/abs/2018A&A...619A..33R} {619, A33}

\bibitem[\protect\citeauthoryear{{Romanova} \& {Owocki}}{{Romanova} \&
  {Owocki}}{2016}]{2016smfu.book..347R}
{Romanova} M.~M.,  {Owocki} S.~P.,  2016, in {Beskin} V.~S.,  {Balogh} A.,
  {Falanga} M.,  {Lyutikov} M.,  {Mereghetti} S.,  {Piran} T.,   {Treumann}
  R.~A.,  eds,  Space Sciences Series of ISSI Vol. 54, The Strongest Magnetic
  Fields in the Universe. Springer, p.~347,
  \mn@doi{10.1007/978-1-4939-3550-5_11}

\bibitem[\protect\citeauthoryear{{Sadakane} \& {Nishimura}}{{Sadakane} \&
  {Nishimura}}{2017}]{2017PASJ...69...48S}
{Sadakane} K.,  {Nishimura} M.,  2017, \mn@doi [\pasj] {10.1093/pasj/psx024},
  \href {https://ui.adsabs.harvard.edu/abs/2017PASJ...69...48S} {69, 48}

\bibitem[\protect\citeauthoryear{{Shultz}}{{Shultz}}{2020}]{2020pase.conf...54S}
{Shultz} M.~E.,  2020, in {Wade} G.,  {Alecian} E.,  {Bohlender} D.,   {Sigut}
  A.,  eds,  Proc. Polish Astron. Soc. Vol. 11, Stellar Magnetism: A Workshop
  in Honour of the Career and Contributions of John D. Landstreet. pp 54--65
  (\mn@eprint {arXiv} {1912.08280})

\bibitem[\protect\citeauthoryear{{Shultz} et~al.,}{{Shultz}
  et~al.}{2018}]{2018MNRAS.475.5144S}
{Shultz} M.~E.,  et~al., 2018, \mn@doi [\mnras] {10.1093/mnras/sty103}, \href
  {https://ui.adsabs.harvard.edu/abs/2018MNRAS.475.5144S} {475, 5144}

\bibitem[\protect\citeauthoryear{{Shultz} et~al.,}{{Shultz}
  et~al.}{2022}]{2022MNRAS.513.1429S}
{Shultz} M.~E.,  et~al., 2022, \mn@doi [\mnras] {10.1093/mnras/stac136}, \href
  {https://ui.adsabs.harvard.edu/abs/2022MNRAS.513.1429S} {513, 1429}

\bibitem[\protect\citeauthoryear{{Soret}, {G{\'e}rard}, {Libert},
  {Shematovich}, {Bisikalo}, {Stiepen}  \& {Bertaux}}{{Soret}
  et~al.}{2016}]{2016Icar..264..398S}
{Soret} L.,  {G{\'e}rard} J.-C.,  {Libert} L.,  {Shematovich} V.~I.,
  {Bisikalo} D.~V.,  {Stiepen} A.,   {Bertaux} J.-L.,  2016, \mn@doi [\icarus]
  {10.1016/j.icarus.2015.09.023}, \href
  {https://ui.adsabs.harvard.edu/abs/2016Icar..264..398S} {264, 398}

\bibitem[\protect\citeauthoryear{{Townsend} \& {Owocki}}{{Townsend} \&
  {Owocki}}{2005}]{2005MNRAS.357..251T}
{Townsend} R.~H.~D.,  {Owocki} S.~P.,  2005, \mn@doi [\mnras]
  {10.1111/j.1365-2966.2005.08642.x}, \href
  {https://ui.adsabs.harvard.edu/abs/2005MNRAS.357..251T} {357, 251}

\bibitem[\protect\citeauthoryear{{Trigilio}, {Leto}, {Umana}, {Leone}  \&
  {Buemi}}{{Trigilio} et~al.}{2004}]{2004A&A...418..593T}
{Trigilio} C.,  {Leto} P.,  {Umana} G.,  {Leone} F.,   {Buemi} C.~S.,  2004,
  \mn@doi [\aap] {10.1051/0004-6361:20040060}, \href
  {https://ui.adsabs.harvard.edu/abs/2004A&A...418..593T} {418, 593}

\bibitem[\protect\citeauthoryear{{Trigilio}, {Leto}, {Umana}, {Buemi}  \&
  {Leone}}{{Trigilio} et~al.}{2011}]{2011ApJ...739L..10T}
{Trigilio} C.,  {Leto} P.,  {Umana} G.,  {Buemi} C.~S.,   {Leone} F.,  2011,
  \mn@doi [\apjl] {10.1088/2041-8205/739/1/L10}, \href
  {https://ui.adsabs.harvard.edu/abs/2011ApJ...739L..10T} {739, L10}

\bibitem[\protect\citeauthoryear{{Vidotto}, {Feeney}  \& {Groh}}{{Vidotto}
  et~al.}{2019}]{2019MNRAS.488..633V}
{Vidotto} A.~A.,  {Feeney} N.,   {Groh} J.~H.,  2019, \mn@doi [\mnras]
  {10.1093/mnras/stz1696}, \href
  {https://ui.adsabs.harvard.edu/abs/2019MNRAS.488..633V} {488, 633}

\bibitem[\protect\citeauthoryear{{Vu{\v{c}}kovi{\'c}}, {{\O}stensen},
  {N{\'e}meth}, {Bloemen}  \& {P{\'a}pics}}{{Vu{\v{c}}kovi{\'c}}
  et~al.}{2016}]{2016A&A...586A.146V}
{Vu{\v{c}}kovi{\'c}} M.,  {{\O}stensen} R.~H.,  {N{\'e}meth} P.,  {Bloemen} S.,
    {P{\'a}pics} P.~I.,  2016, \mn@doi [\aap] {10.1051/0004-6361/201526552},
  \href {https://ui.adsabs.harvard.edu/abs/2016A&A...586A.146V} {586, A146}

\bibitem[\protect\citeauthoryear{{Yakunin}, {Romanyuk}, {Mikul{\'a}{\v{s}}ek},
  {Jan{\'\i}k}, {Moiseeva}  \& {H{\"u}mmerich}}{{Yakunin}
  et~al.}{2020}]{2020AzAJ...15a..93Y}
{Yakunin} I.~A.,  {Romanyuk} I.~I.,  {Mikul{\'a}{\v{s}}ek} Z.,  {Jan{\'\i}k}
  J.,  {Moiseeva} A.,   {H{\"u}mmerich} S.,  2020, Azerbaijani Astronomical
  Journal, \href {https://ui.adsabs.harvard.edu/abs/2020AzAJ...15a..93Y} {15,
  93}

\bibitem[\protect\citeauthoryear{{ud-Doula} \& {Naz{\'e}}}{{ud-Doula} \&
  {Naz{\'e}}}{2016}]{2016AdSpR..58..680U}
{ud-Doula} A.,  {Naz{\'e}} Y.,  2016, \mn@doi [Advances in Space Research]
  {10.1016/j.asr.2015.09.025}, \href
  {https://ui.adsabs.harvard.edu/abs/2016AdSpR..58..680U} {58, 680}

\bibitem[\protect\citeauthoryear{{ud-Doula} \& {Owocki}}{{ud-Doula} \&
  {Owocki}}{2002}]{2002ApJ...576..413U}
{ud-Doula} A.,  {Owocki} S.~P.,  2002, \mn@doi [\apj] {10.1086/341543}, \href
  {https://ui.adsabs.harvard.edu/abs/2002ApJ...576..413U} {576, 413}

\bibitem[\protect\citeauthoryear{{ud-Doula} \& {Owocki}}{{ud-Doula} \&
  {Owocki}}{2022}]{2022hxga.book...46U}
{ud-Doula} A.,  {Owocki} S.,  2022, in , Handbook of X-ray and Gamma-ray
  Astrophysics.
Springer, p.~46, \mn@doi{10.1007/978-981-16-4544-0_80-1}

\bibitem[\protect\citeauthoryear{{ud-Doula}, {Townsend}  \&
  {Owocki}}{{ud-Doula} et~al.}{2006}]{udDoula2006}
{ud-Doula} A.,  {Townsend} R. H.~D.,   {Owocki} S.~P.,  2006, \mn@doi [\apjl]
  {10.1086/503382}, \href
  {https://ui.adsabs.harvard.edu/abs/2006ApJ...640L.191U} {640, L191}

\bibitem[\protect\citeauthoryear{{ud-Doula}, {Owocki}  \&
  {Townsend}}{{ud-Doula} et~al.}{2008}]{2008MNRAS.385...97U}
{ud-Doula} A.,  {Owocki} S.~P.,   {Townsend} R. H.~D.,  2008, \mn@doi [\mnras]
  {10.1111/j.1365-2966.2008.12840.x}, \href
  {https://ui.adsabs.harvard.edu/abs/2008MNRAS.385...97U} {385, 97}

\makeatother
\end{thebibliography}




\appendix

\section{FUSE spectroscopy of magnetic stars}
\label{appendixFUSE}

\begin{figure}
	
    \includegraphics[width=\columnwidth]{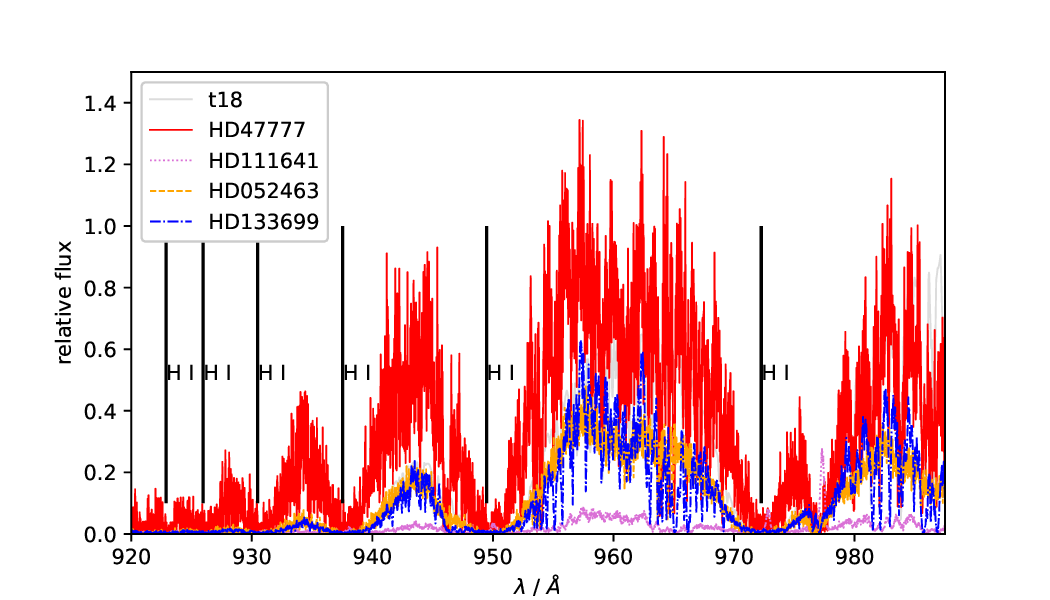}
    \includegraphics[width=\columnwidth]{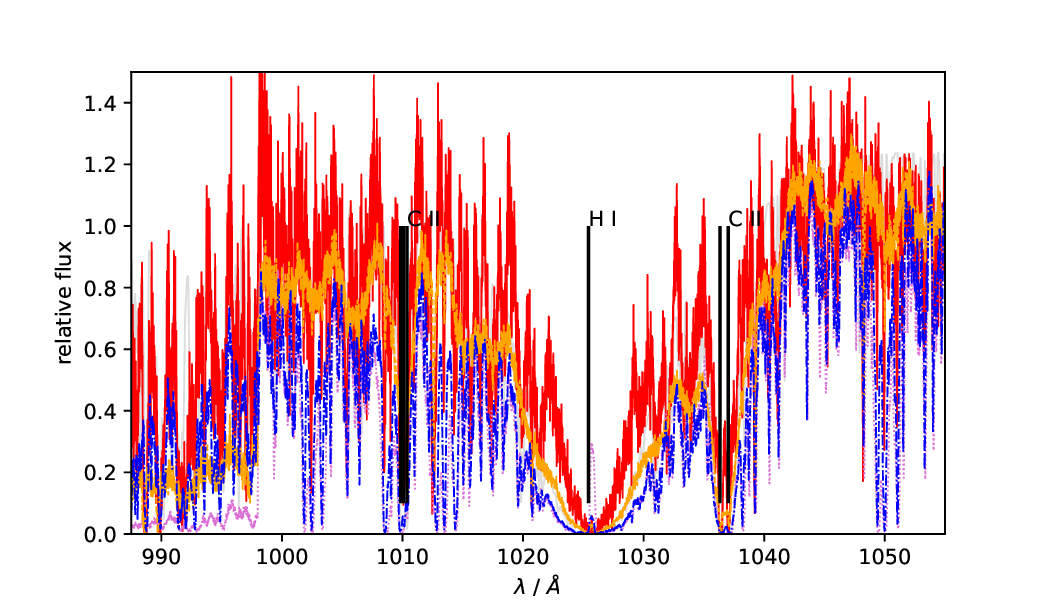}
    \includegraphics[width=\columnwidth]{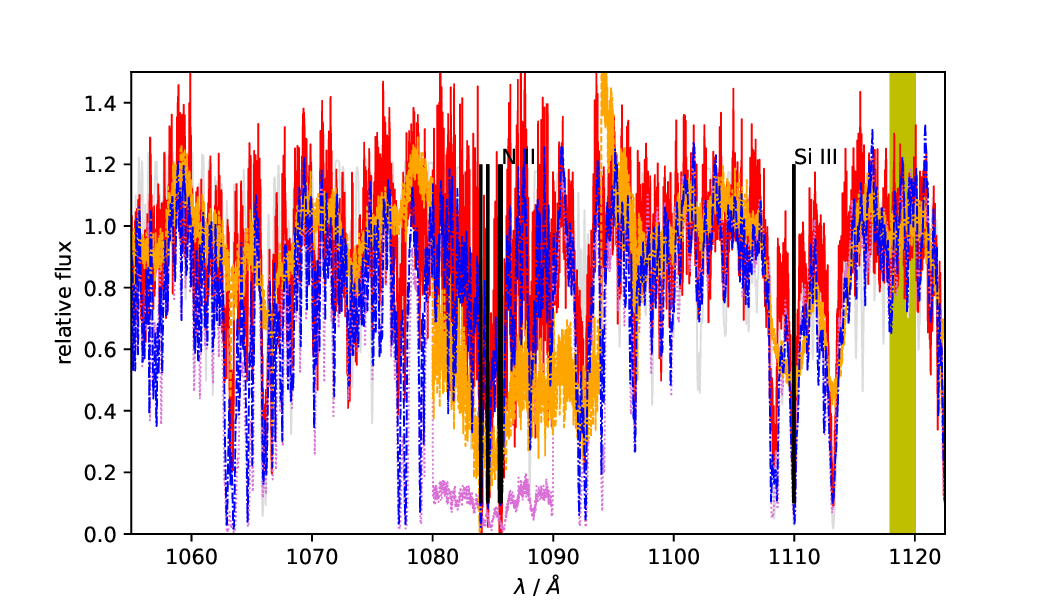}
    \includegraphics[width=\columnwidth]{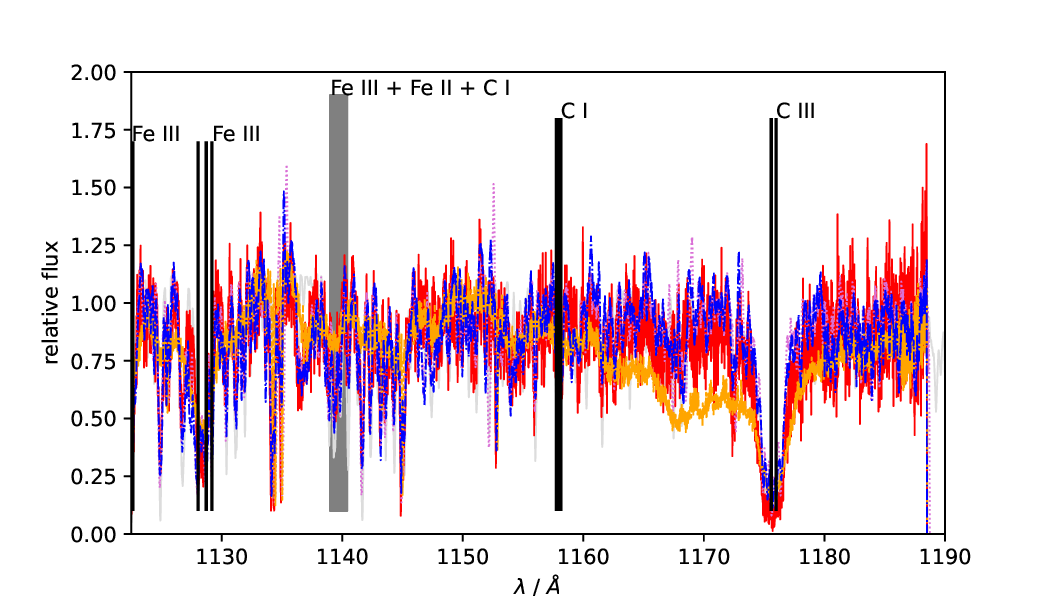}
    \caption{Same as Fig.~\ref{fig:HD200775}, but for magnetic star HD 47777 and t18 model.}
    \label{fig:HD47777}
\end{figure}

\begin{figure}
	
    \includegraphics[width=\columnwidth]{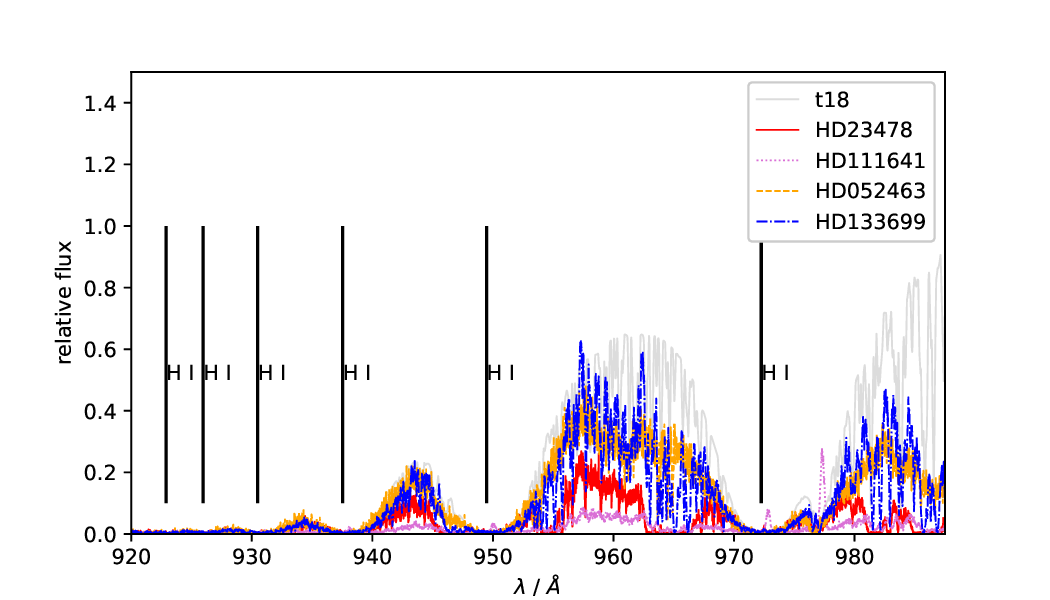}
    \includegraphics[width=\columnwidth]{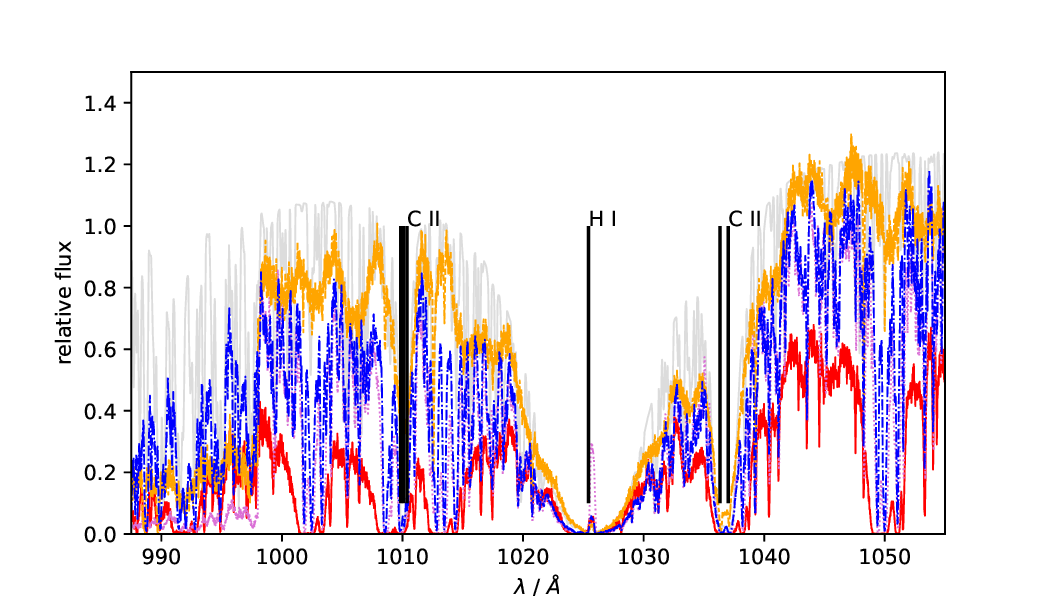}
    \includegraphics[width=\columnwidth]{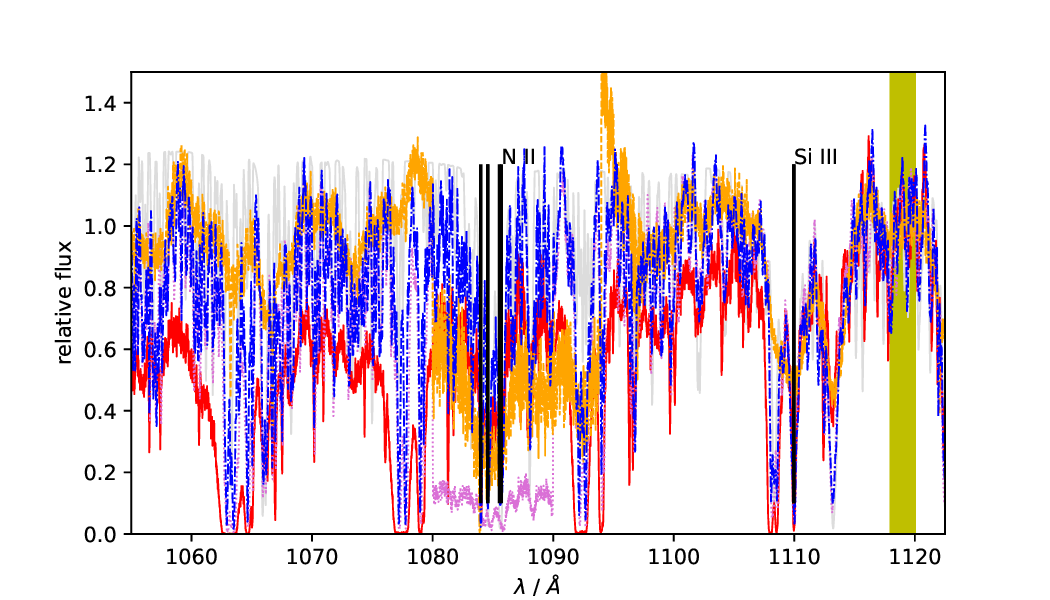}
    \includegraphics[width=\columnwidth]{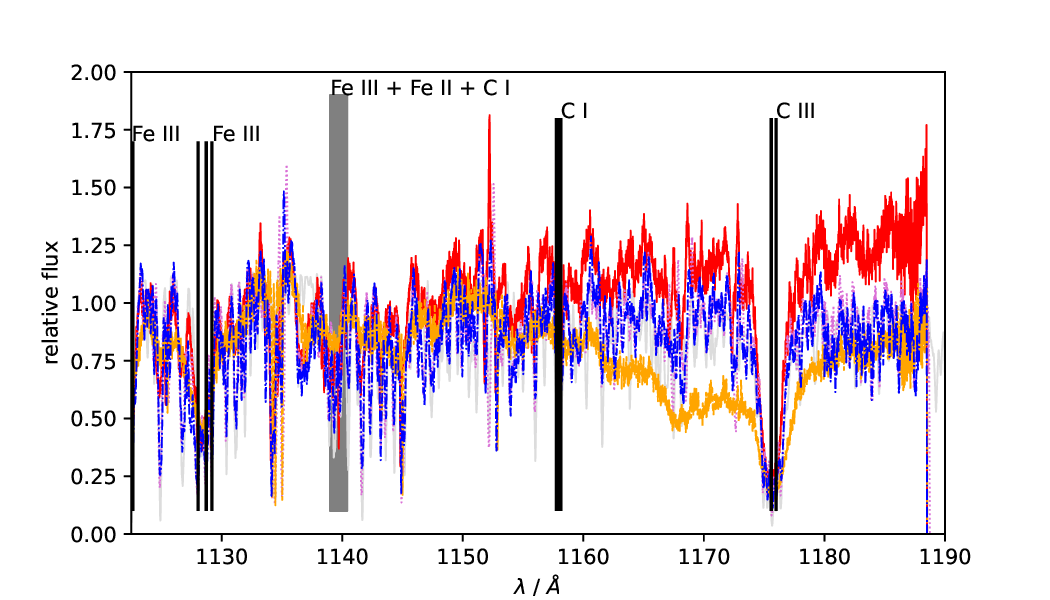}
    \caption{Same as Fig.~\ref{fig:HD200775}, but for magnetic star HD 23478 and t18 model}
    \label{fig:HD23478}
\end{figure}

\begin{figure}
    \includegraphics[width=\columnwidth]{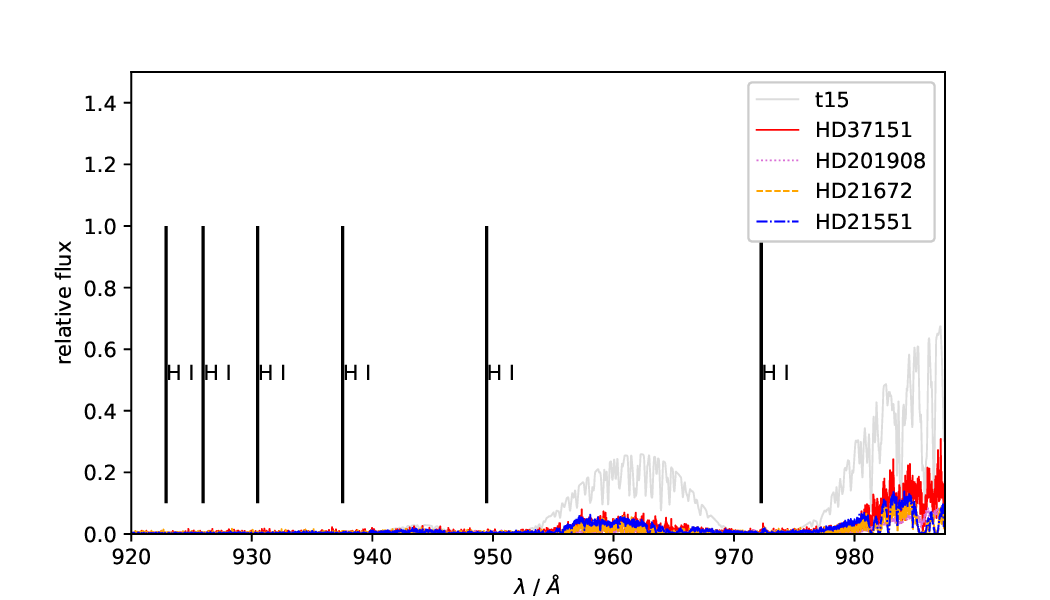}
    \includegraphics[width=\columnwidth]{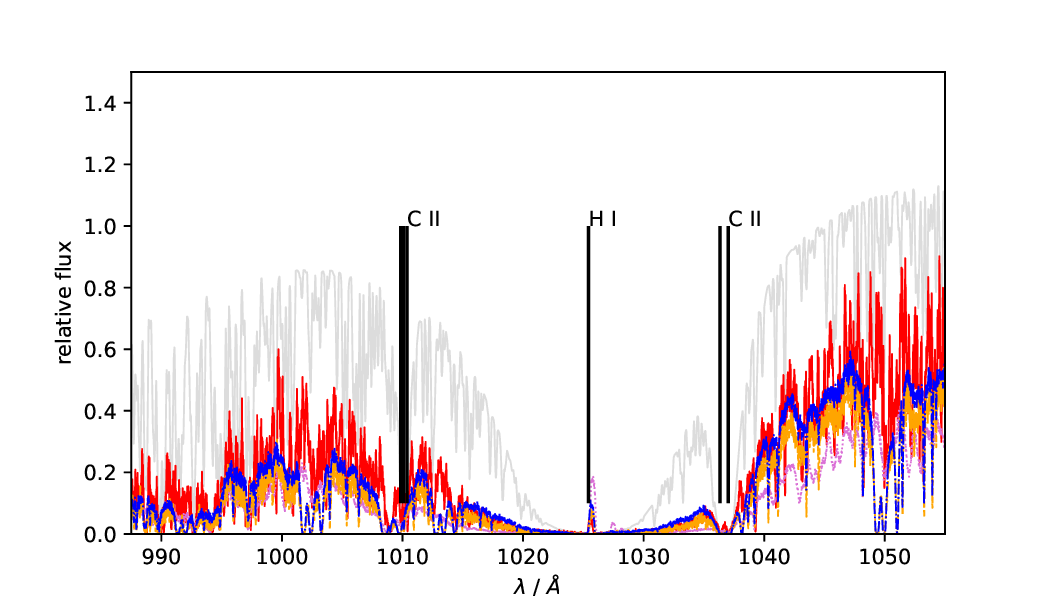}
    \includegraphics[width=\columnwidth]{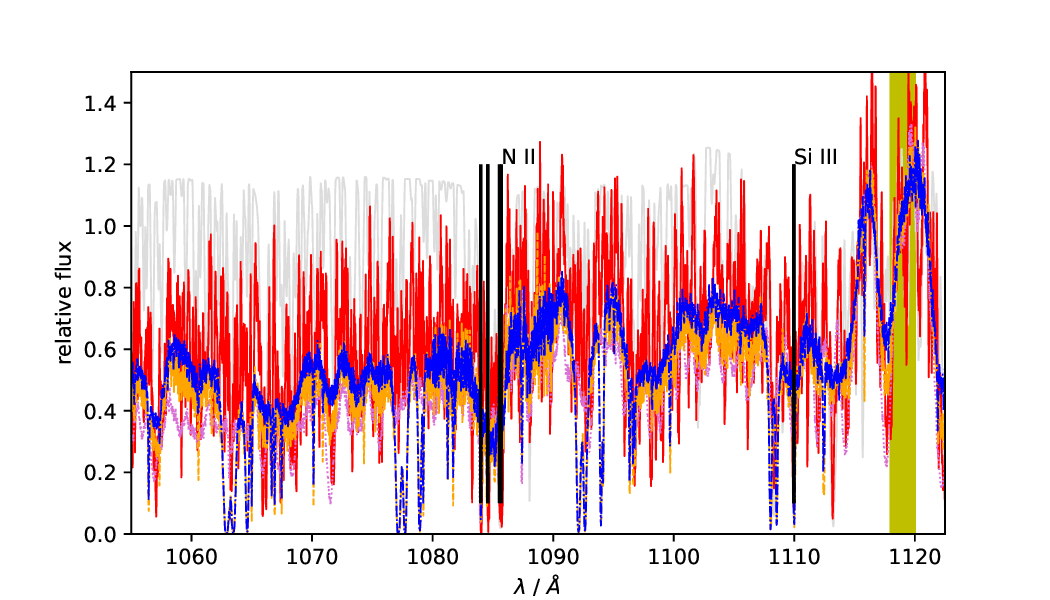}
    \includegraphics[width=\columnwidth]{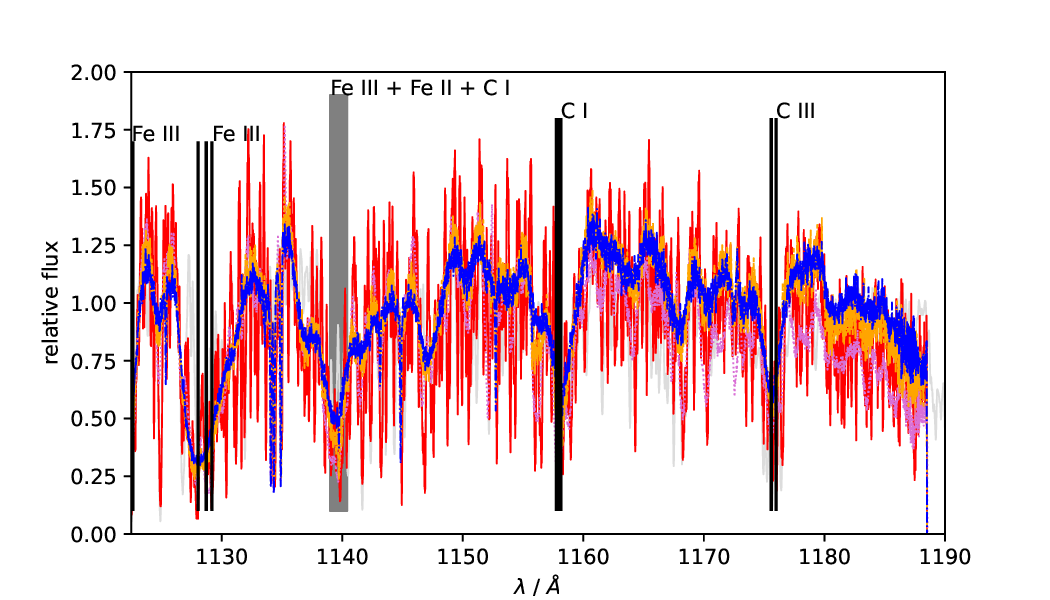}
    \caption{Same as Fig.~\ref{fig:HD200775}, but for magnetic star HD 37151 and t15 model}
    \label{fig:HD37151}
\end{figure}

\begin{figure}
    \includegraphics[width=\columnwidth]{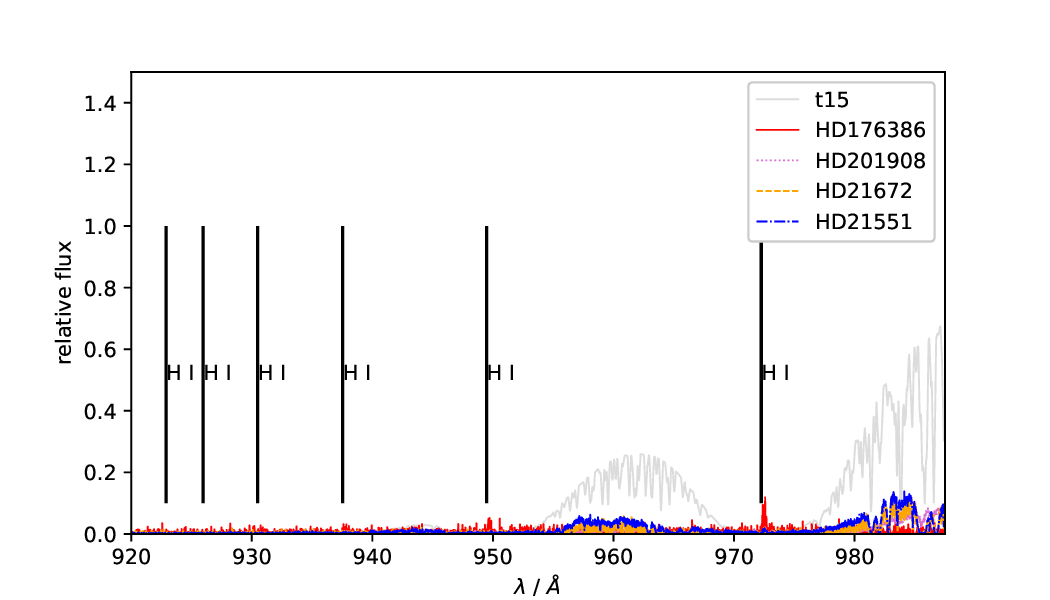}
    \includegraphics[width=\columnwidth]{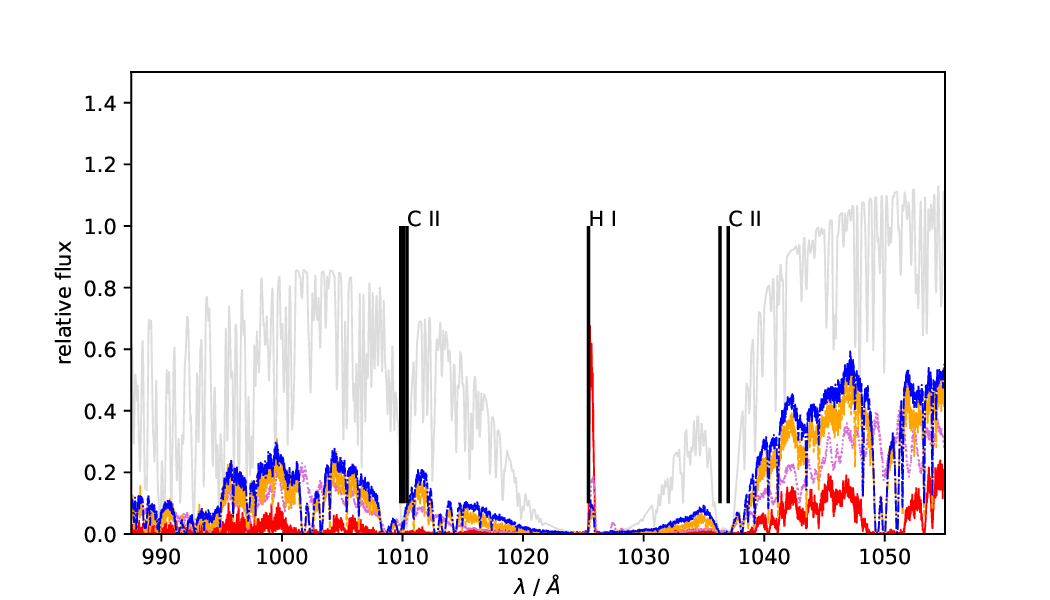}
    \includegraphics[width=\columnwidth]{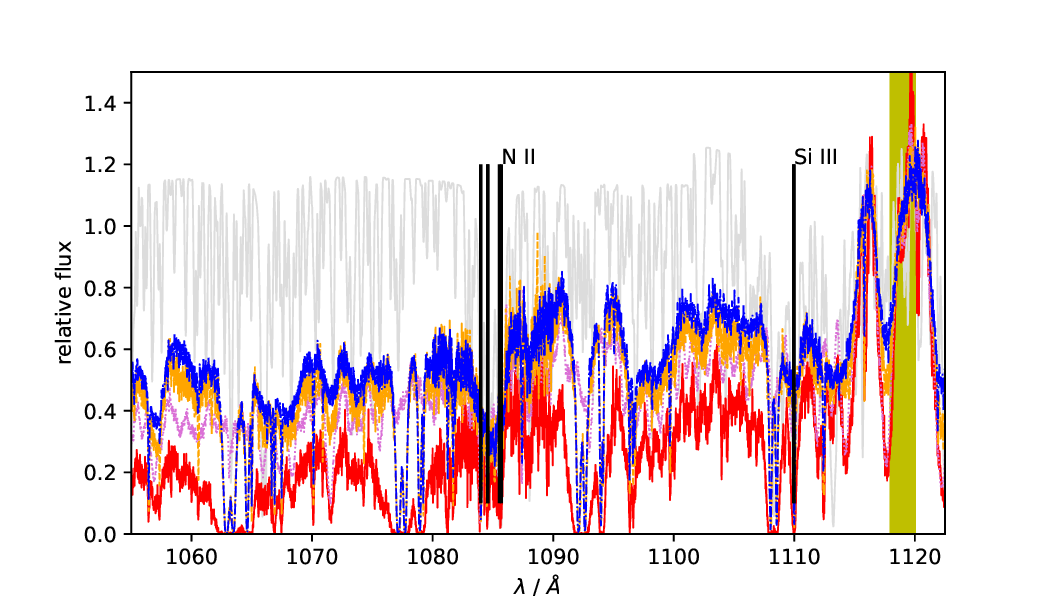}
    \includegraphics[width=\columnwidth]{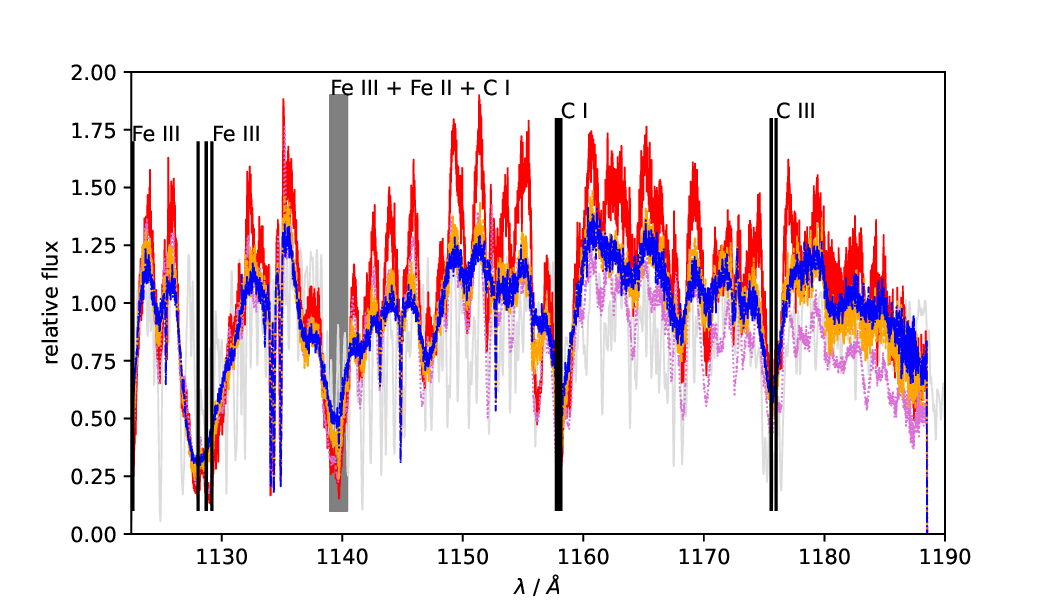}
    \caption{Same as Fig.~\ref{fig:HD200775}, but for magnetic star HD 176386 and t15 model}
    \label{fig:HD176386}
\end{figure}

\begin{figure}
    \includegraphics[width=\columnwidth]{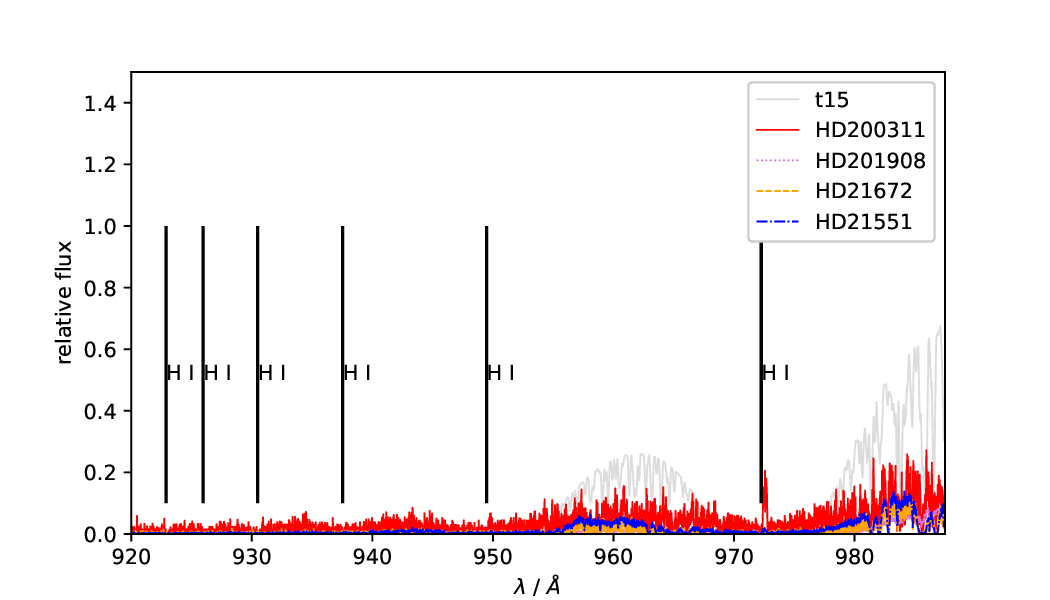}
    \includegraphics[width=\columnwidth]{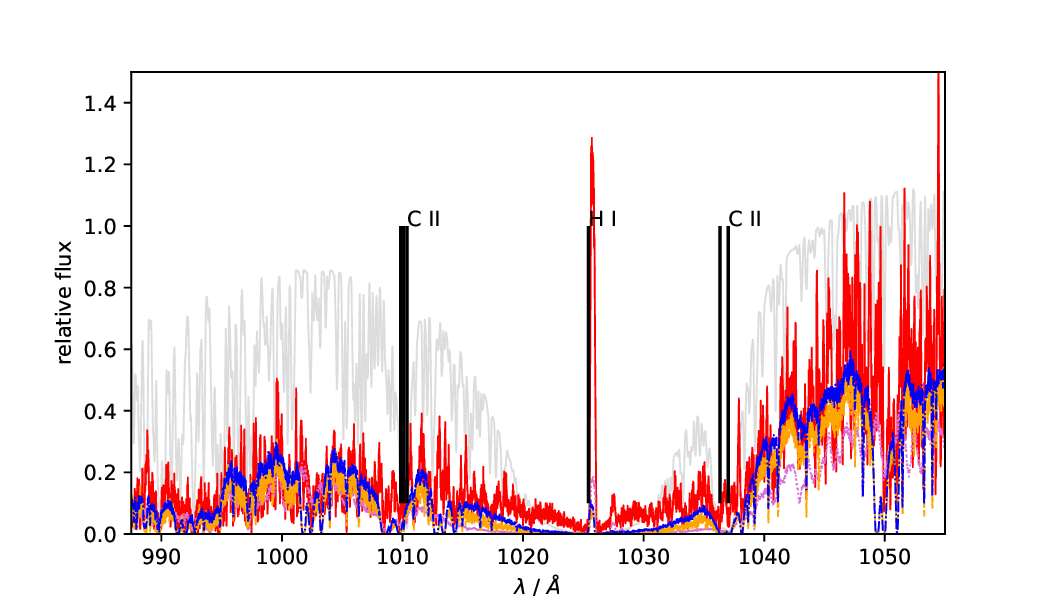}
    \includegraphics[width=\columnwidth]{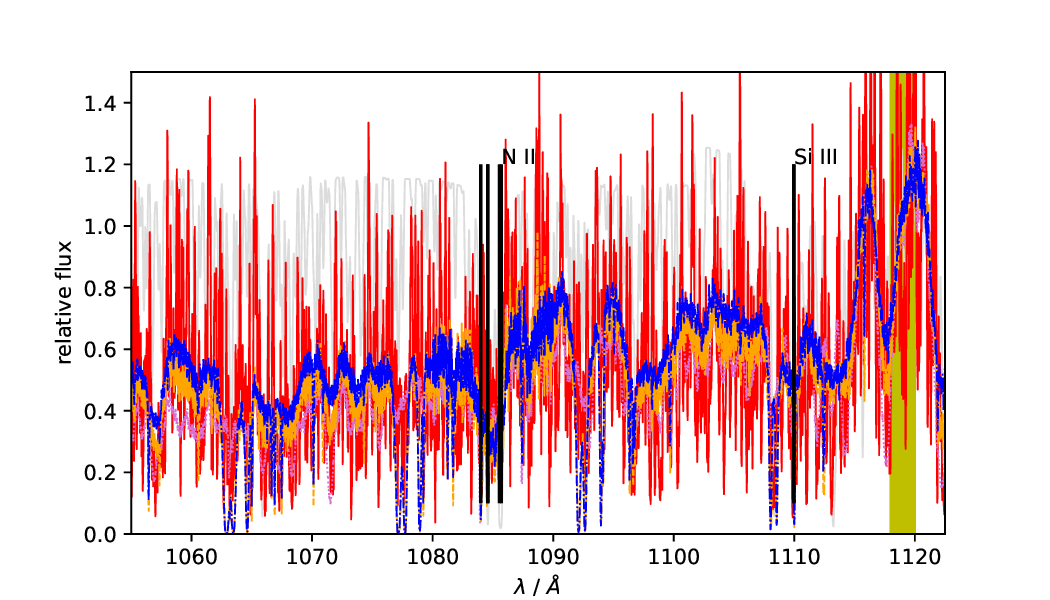}
    \includegraphics[width=\columnwidth]{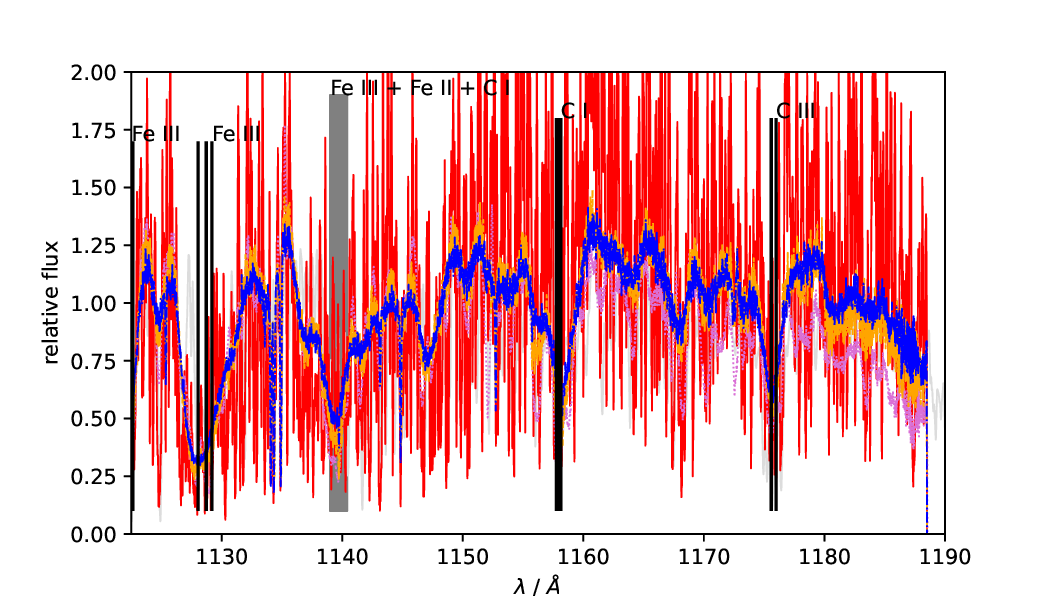}
    \caption{Same as Fig.~\ref{fig:HD200775}, but for magnetic star HD 200311 and t15 model}
    \label{fig:HD200311}
\end{figure}

 \begin{table}
    \centering
    \caption{Reference nonmagnetic stars observed with FUSE.}
    \begin{tabular}{l|c|c}
    \hline
    ID & Sp. Type (Simbad) & Used as reference for \\
    \hline
        HD 52463 & B3V & HD 23478, HD 47777, HD  200775 \\
        HD 111641 & B3V & HD 23478, HD 47777, HD  200775  \\
        HD 133699 & B3V & HD 23478, HD 47777, HD  200775   \\
        HD 201908 & B8V & HD 37151, HD 176386, HD 200311\\
        HD 21551 & B8V & HD 37151, HD 176386, HD 200311\\
        HD 21672 & B8V & HD 37151, HD 176386, HD 200311 \\
    \hline
    \end{tabular}

    \label{tab:nonBstar}
\end{table}
\bsp	
\label{lastpage}
\end{document}